\documentclass[aps,pra,longbibliography,preprint]{revtex4-2}

\usepackage{graphicx}
\usepackage{dcolumn}
\usepackage{bm}
\usepackage{amsmath}
\usepackage{amssymb}

\usepackage{color}
\usepackage[colorlinks,bookmarks=false,citecolor=darkblue,linkcolor=red,urlcolor=blue]{hyperref}

\definecolor{darkred}{rgb}{0.7,0.0,0.0}
\definecolor{darkblue}{rgb}{0,0.02,0.45}

\newcommand{\KBa}{KBaYb(BO$_3$)$_2$}

\graphicspath{{figures/}}

\begin{document}

\title{Frustrated magnet for adiabatic demagnetization cooling to milli-Kelvin temperatures}


\author{Y. Tokiwa*$^{1,2}$}
\author{S. Bachus$^1$}
\author{K. Kavita$^1$}
\author{A. Jesche$^1$}
\author{A. A. Tsirlin$^1$}
\author{P. Gegenwart$^1$}
\affiliation{Experimental Physics VI, Center for Electronic Correlations and Magnetism, University of Augsburg, Germany}
\affiliation{Advanced Science Research Center, Japan Atomic Energy Agency, Tokai, Naka, Ibaraki, Japan}
\affiliation{*ytokiwa@gmail.com}
\date{\today}

\begin{abstract}

Generation of very low temperatures has been crucially important for applications and fundamental research, as low-temperature quantum coherence enables operation of quantum computers and formation of exotic quantum states, such as superfluidity and superconductivity. One of the major techniques to reach milli-Kelvin temperatures is adiabatic demagnetization refrigeration (ADR). This method uses almost non-interacting magnetic moments of paramagnetic salts where large distances suppress interactions between the magnetic ions. The large spatial separations are facilitated by water molecules, with a drawback of reduced stability of the material. Here, we show that an H$_2$O-free frustrated magnet KBaYb(BO$_3$)$_2$ can be ideal refrigerant for ADR, achieving at least 22\,mK upon demagnetization under adiabatic conditions. Compared to conventional refrigerants, KBaYb(BO$_3)_2$ does not degrade even under high temperatures and ultra-high vacuum conditions. Further, its frustrated magnetic network and structural randomness enable cooling to temperatures several times lower than the energy scale of magnetic interactions, which is the main limiting factor for the base temperature of conventional refrigerants.

\end{abstract}

\maketitle


\section{Introduction}

Suppression of thermal fluctuations by lowering temperature gives access to intricate and potentially usable quantum effects. Major discoveries, such as quantum Hall effect and superfluidity/superconductivity, have been made by explorations of matter close to absolute zero~\cite{RevModPhys.83.1193,SC}. Recently, the development of quantum computers and sensors for dark matter detection rendered low-temperature refrigeration an important technological challenge~\cite{Zhu2011,LTD}. One of the viable methods for reaching milli-K (mK) temperature range is adiabatic demagnetization refrigeration (ADR) using paramagnetic salts~\cite{Pobell-92,debye,giauque}. Its main advantages over the currently dominant technique, $^3$He-$^4$He dilution refrigeration, are the simple construction of a cooling device and its operation without the usage of expensive $^3$He. The recent crisis of $^3$He, which was caused by the increased demand due to construction of neutron detectors for defense against nuclear terrorism, raised serious concerns about the strong dependence of the current technology on such a scarcely available and ever more expensive gas~\cite{Shea-CRS10,Kouzes,Cho-Science2009}. This triggered renewed interest in ADR, as well as interesting proposals of completely new types of ADR materials~\cite{Wolf-PNAS11,Jang-NC15,YbGG,Tokiwae1600835,Evangelisti,Baniodeh,Zhitomirsky,Er227,Hu_2008}.

The only advantage of $^3$He-$^4$He dilution refrigeration is its capability of continuous cooling, while conventional ADR is a single-shot technique. This makes the $^3$He-$^4$He dilution refrigeration more commonly used than ADR. However, the situation may change thanks to recent developments of continuous ADR cooling~\cite{SHIRRON2004581,BARTLETT2010582} and availability of commercial continuous refrigerators based on ADR~\cite{kiutra}. Therefore, ADR has the potential of becoming the main cooling technology already in near future, at least in the mK temperature range.

\begin{figure}
\includegraphics[width=0.8\linewidth,keepaspectratio]{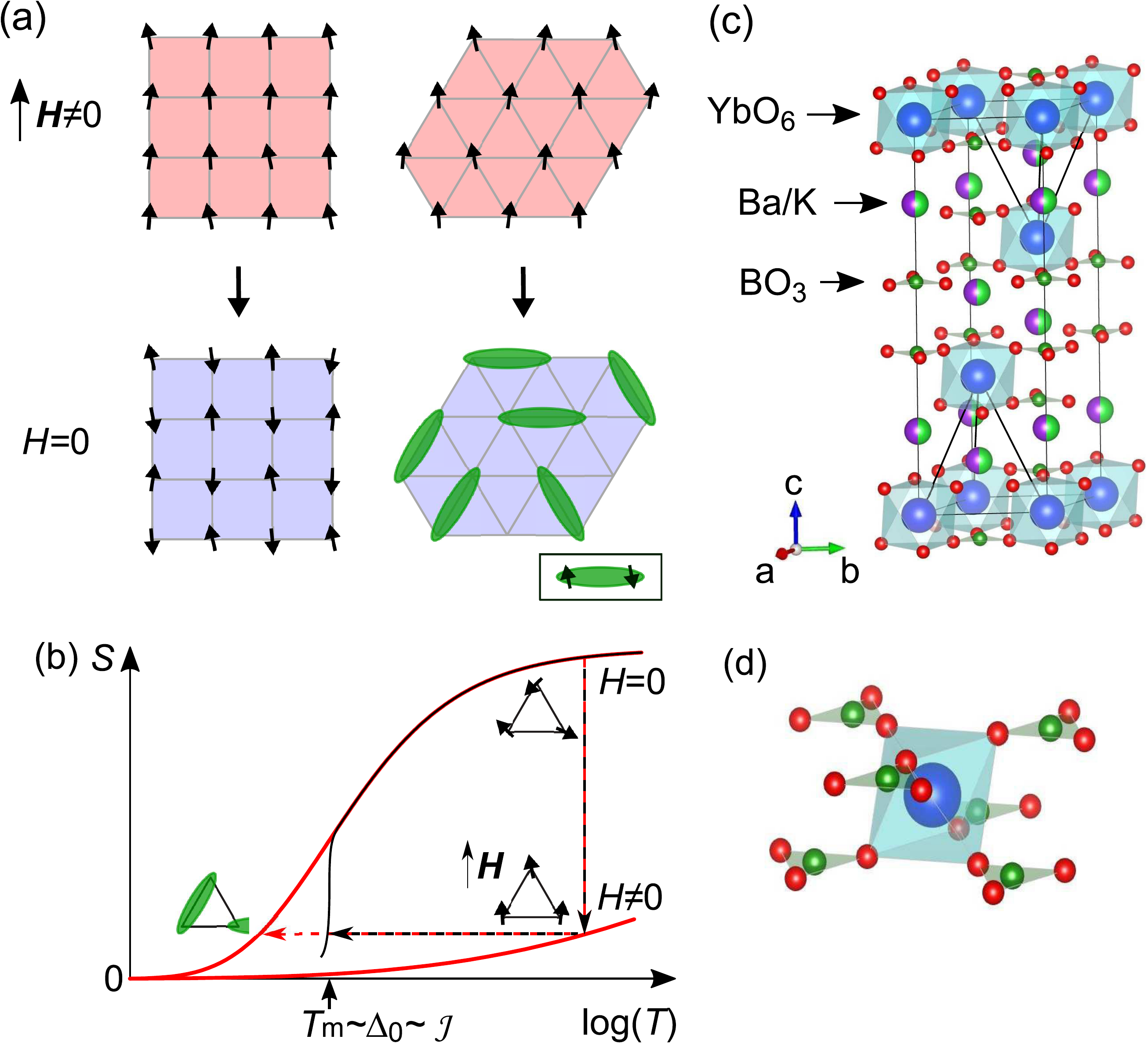}%
\caption{\label{Structure}(a) Adiabatic demagnetization processes of a conventional paramagnet (left) and a frustrated magnet with impeded magnetic order (right). Green ovals represent singlet pairings in a short-range correlated but long-range disordered state, such as a spin liquid caused by magnetic frustration~\cite{balents-nature10,li2020} or a random-singlet state caused by structural disorder~\cite{dey2020}. (b) Entropy curves at zero and a finite magnetic field as a function of temperature for conventional paramagnets (black) and a system with suppressed magnetic order (red). While the former orders below some temperature $T_{\rm m}$ around $\Delta_0$ or $\mathcal J$, the latter remains disordered down to much lower temperatures. For simplicity, a Schottky-type increase in the entropy is assumed. Note however that the entropy increase may be much slower in a spin liquid because of its broad energy spectrum. $\Delta_0$ is an energy level splitting at zero external field because of the internal field caused by magnetic interactions. Arrows with red and black dotted lines represent cooling processes for the conventional paramagnets (black) and frustrated magnets (red). (c) Crystal structure of KBaYb(BO$_3$)$_2$ with triangular layers of the Yb$^{3+}$ ions. Black solid lines highlight relative displacement of the adjacent triangular layers. (d) YbO$_6$ octahedron surrounded by the BO$_3$ triangles. }
\end{figure}

ADR uses magnetic moments of almost-ideal paramagnets with very weak magnetic interaction $\mathcal J$. Because the interaction is weak, magnetic moments are easily aligned by external magnetic field, causing reduction of entropy (Figure~\ref{Structure}(a,b)). Even at zero external magnetic field $H=0$, magnetic moments of such almost-ideal paramagnets experience some small internal magnetic field produced by adjacent magnetic moments through magnetic interactions. This causes a tiny Zeeman splitting and magnetic order at the same energy scale ($\sim\Delta_0\sim\mathcal J$). Therefore, entropy, which is the driving force of ADR, decreases to zero below the temperature of $T\sim\mathcal J$, thus putting a limit on the end temperature $T_{\rm f}\sim\mathcal J$ that can be reached via ADR with this material (Figure~\ref{Structure}(b)). As indicated by a black horizontal arrow in Figure~\ref{Structure}(b), entropy difference between $H$=0 and $H$$\neq$0 is the key for ADR. While a perfect paramagnet with zero magnetic interactions would be an ideal refrigerant, having maximum entropy at zero field down to zero temperature, there are always weak but finite interactions in real materials.

In commonly used paramagnetic salts, these interactions are reduced by large distances between the magnetic ions, which are separated by water molecules. However, abundance of water makes these salts prone to decomposition. They deliquesce in humid atmosphere and dehydrate in vacuum or upon even mild heating. Therefore, for repeated use without degradation, stable water-free materials with very weak magnetic interactions are desirable. Furthermore, ADR would be certainly beneficial for applications in ultra-high-vacuum (UHV) apparatus, especially in scanning tunneling microscopy and angle-resolved photoemission spectroscopy where the necessity of chamber baking at high temperature and high vacuum for reaching UHV makes the use of current ADR materials essentially impossible.

For many decades ever since P. Debye~\cite{debye} and W.F. Giauque~\cite{giauque} independently proposed ADR, water-containing paramagnetic salts have been materials of choice for cooling in the mK range~\cite{Pobell-92,cpa,Viches-PR66}. Here, we demonstrate the effective refrigeration with an H$_2$O-free frustrated magnet \KBa{} that shows excellent ADR performance on par with conventional water-containing paramagnetic salts, and achieves an end temperature $T_{\rm f}$ of at least 22\,mK, starting from a temperature of 2\,K at a magnetic field of 5\,T. In this compound, Yb$^{3+}$ ions are responsible for magnetic properties and thus for the ADR. At sufficiently low temperatures, only the lowest Kramers doublet of Yb$^{3+}$ is occupied~\cite{li2020}.

The crystal structure of \KBa{} depicted in Figs.~\ref{Structure}(c,d) reveals isolated YbO$_6$ octahedra surrounded by six BO$_3$ triangles. These negatively charged units are interspersed by the positively charged K$^+$ and Ba$^{2+}$ ions that occupy the single crystallographic site. The shorter Yb--Yb distances of 5.41\,\r A within the $ab$ plane connect magnetic Yb-ions into a triangular lattice, the feature that already caused interest in this~\cite{sanders2017,guo2019a} and isostructural~\cite{guo2019b,guo2019c,guo2019d} rare-earth borates from the perspective of frustrated magnetism~\cite{li2020}. The interplane Yb-Yb distances are only slightly longer (6.64\,\r A) and cause additional frustration, because adjacent triangular layers are shifted relative to each other, as shown by solid lines in Fig. 1(c), connecting Yb-ions from the neighboring layers. Such geometrical frustration may shift the magnetic ordering transition toward lower temperatures or even suppress the magnetic order completely. The distribution of K$^+$ and Ba$^{2+}$ results in uneven electric fields acting on the magnetic Yb$^{3+}$ ions and may lead to random exchange couplings~\cite{li2020} that are also detrimental for the magnetic ordering. Sizable remaining entropy due to the suppression of the magnetic long-range order facilitates the cooling even below $\sim\mathcal J$ (Figure~\ref{Structure}(b)). Moreover, the presence of soft modes in frustrated magnets amplifies magnetocaloric effect and further lowers the end temperature of the cooling process~\cite{Zhitomirsky}. The enhanced ADR performance was indeed observed in frustrated magnets, but temperatures in the mK-range could not be reached because of the relatively strong magnetic interactions~\cite{YbGG,GTO}. \KBa{} is a novel type of an H$_2$O-free refrigerant combining several properties that are beneficial for reaching lowest temperatures: large separation between the magnetic ions, geometrical frustration of residual magnetic interactions, as well as their potential randomness.

\section{Results}

\begin{figure}
\includegraphics[width=\linewidth,keepaspectratio]{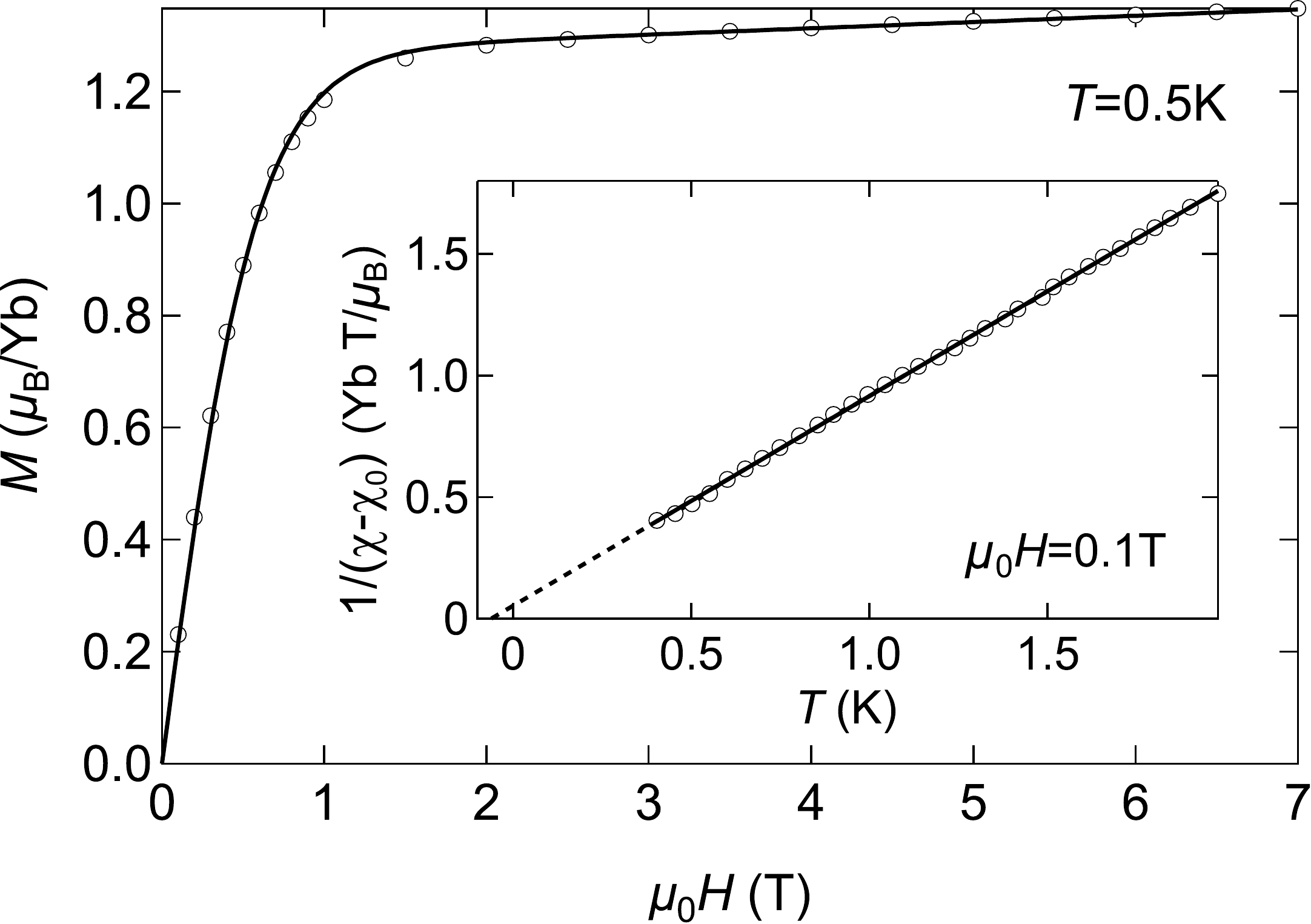}%
\caption{\label{Magnetic_Properties} Field-dependent magnetization of \KBa{} at 0.5\,K. Solid line is a fit to the data with $gJB_J(x)$+$\chi_0H$ where $g$ is the $g$-factor, $J$=$\frac12$ for the lowest Kramers doublet of Yb$^{3+}$, and $B_J$ is the Brillouin function for $J$=$\frac12$ with $x$=$(g\mu_{\rm B}J\mu_0H)/(k_{\rm B}T)$. The second term is due to the van Vleck susceptibility $\chi_0$. The fit returns $g$=2.54 and $\chi_0/\mu_0$=0.0111\,$\mu_{\rm B}$/T. The inset shows inverse magnetic susceptibility, 1/($\chi-\chi_0$) where $\chi$=$\mu_0H$/$M$, measured at 0.1\,T. Solid line is a linear fit in the 0.4\,K$\leq$$T$$\leq$2.0\,K range, resulting in the effective magnetic moment $\mu_{\rm eff}$=2.28\,$\mu_B$ and Curie-Weiss temperature $\theta=-60\pm{2}$\,mK.}
\end{figure}

Magnetization measured down to 0.5\,K shows the typical paramagnetic behavior (Figure~\ref{Magnetic_Properties}). Field-dependent magnetization is perfectly described by the Brillouin function augmented by the van Vleck term $\chi_0H$ that originates from crystal-field excitations. Upon subtracting this van Vleck contribution $\chi_0$ from temperature-dependent susceptibility, one recovers linear behavior in $1/(\chi-\chi_0)$ corresponding to the Curie-Weiss law. The fit returns the effective magnetic moment $\mu_{\rm eff}=2.28$\,$\mu_B$ and Curie-Weiss temperature $\theta=-60\pm{2}$\,mK. We note that the reduced effective magnetic moment from free Yb$^{3+}$ is due to crystalline electric field effect. We show in Supplementary information that at high temperatures $\mu_{\rm eff}$ recovers the free-ion value for Yb$^{3+}$. The small value of $\theta$ indicates that temperatures on the order of 10\,mK would be needed to explore cooperative magnetism and frustrated magnetic behavior of \KBa{}, making this material a quite challenging system to study in comparison with YbMgGaO$_4$ and Yb$^{3+}$ delafossites having Curie-Weiss temperatures of several Kelvin~\cite{li2020}. On the other hand, \KBa{} itself can be used to reach these extremely low temperatures, as we show below.

\begin{figure}
\includegraphics[width=0.8\linewidth,keepaspectratio]{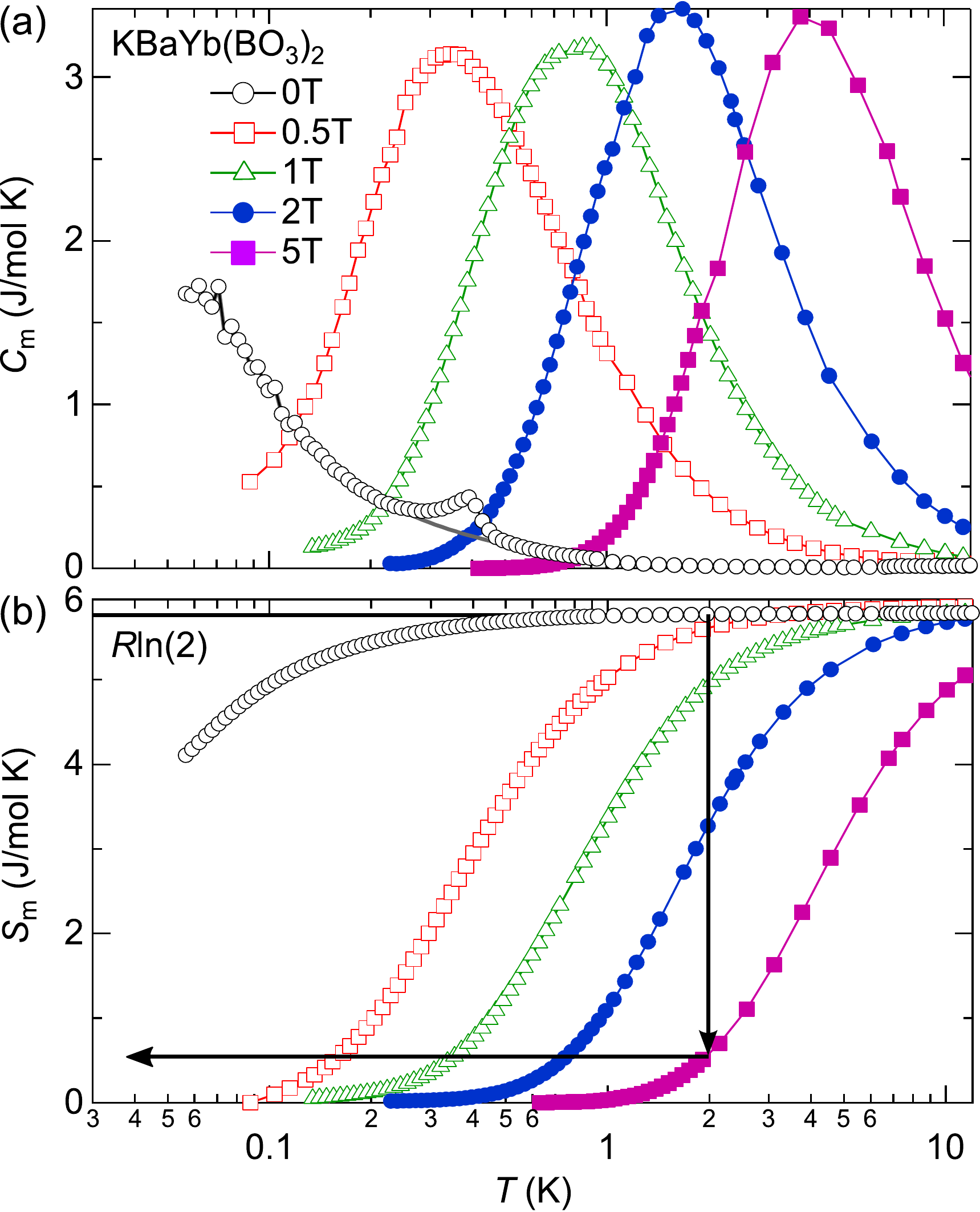}%
\caption{(a) Low-temperature magnetic heat capacity of \KBa{} at several external magnetic fields. Heat capacity of a non-magnetic reference KBaLu(BO$_3$)$_2$ was subtracted from the raw data as a phonon contribution. (b) Magnetic entropy $S_{\rm m}$ calculated by integrating $C_{\rm m}$/$T$. In zero field, the entropy is vertically shifted to match $R\ln 2$ expected for the Kramers doublet. Two arrows show the ADR process, starting from $T=2$\,K at $\mu_{\rm 0}H=5$\,T.  \label{HC_and_Entropy}}
\end{figure}

First hints for the ADR potential of \KBa{} are found in the magnetic specific heat data $C_m(T)$ that were obtained via subtracting the lattice contribution given by the non-magnetic reference compound KBaLu(BO$_3$)$_2$ (Figure~\ref{HC_and_Entropy}a). In zero field, specific heat increases toward lower temperatures and indicates the accumulation of entropy associated with the Kramers doublet of Yb$^{3+}$. Magnetic field splits this doublet into the $j_z=+\frac12$ and $j_z=-\frac12$ levels and causes a Schottky anomaly that shifts toward higher temperatures upon increasing the field. A weak anomaly is seen in the zero-field data around 400\,mK, but its nature is probably extrinsic, as it contains only a tiny amount of the magnetic entropy (See Supplementary Information). Note that a similar anomaly has been reported for isostructural NaBaYb(BO$_3)_2$~\cite{guo2019b}.

By integrating $C_m(T)/T$, we calculate magnetic entropy (Figure~\ref{HC_and_Entropy}b) and observe that already at 0.5\,T the full entropy of $R\ln 2$ associated with the lowest Kramers doublet of Yb$^{3+}$ can be recovered above 80\,mK. Higher fields shift the entropy toward higher temperatures, so that at 5\,T only 10\% of $R\ln 2$ is released below 2\,K. Therefore, by adiabatic demagnetization of \KBa{} starting from 2\,K and 5\,T, one may expect to reach temperatures as low as 30\,mK, as shown by arrows in Figure~\ref{HC_and_Entropy}(b).

We test this prediction in the actual cooling experiment that was performed with a 4.02g pellet containing equal weights of \KBa{} and silver powder (10 to 50\,$\mu$m particle size). Silver improves thermal conductivity within the pellet of insulating \KBa{}. To improve thermal conductance between the particles, we sintered the pressed pellet at 600$^{\circ}$C. The pellet was mounted on a sapphire-plate sample holder, which is held on a plastic frame by four 50\,$\mu$m thin fishing wires. A resistive RuO$_2$ thermometer is attached to the pellet with glue. To suppress the thermal conduction from the heat bath, thin superconducting filaments are used as wire leads on the resistive thermometer for temperature measurements. This setup was attached to the cold bath in a vacuum chamber (with pressure of $\sim$ 10$^{-6}$\,mbar) of the $^3$He-$^4$He dilution refrigerator.

\begin{figure}
\includegraphics[width=\linewidth,keepaspectratio]{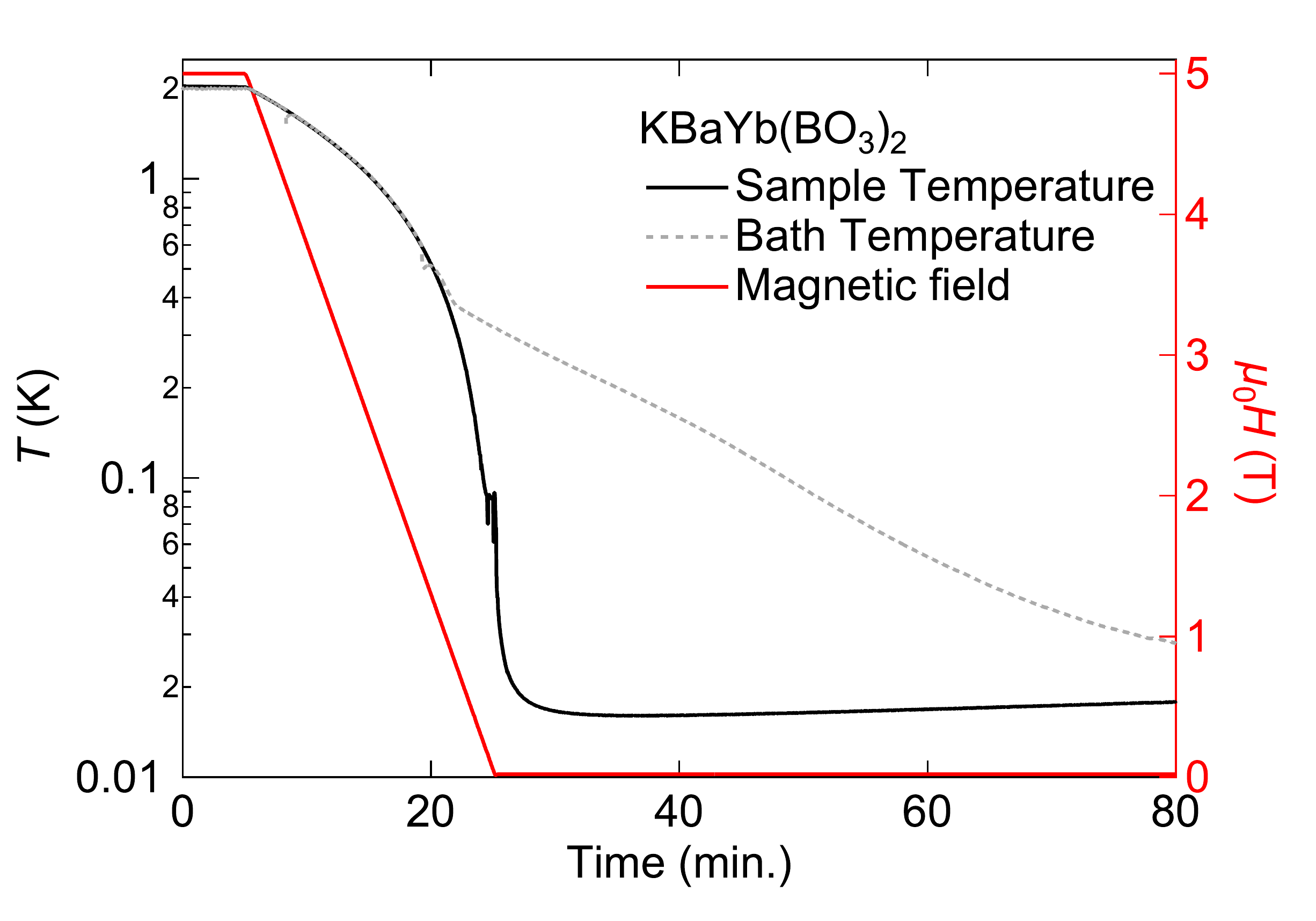}%
\caption{Refrigeration test under nearly adiabatic conditions. The solid black, solid red, and dotted grey lines are sample temperature (left axis), external magnetic field (right axis), and bath temperature (left axis), respectively. The sample pellet has weak thermal connection to the bath. The bath temperature is controlled to follow the sample temperature to minimize the heat flow. Below about 0.4\,K, the bath cannot follow the sample temperature any more, because the cooling rate of ADR is too fast to follow. The sample holder is in a vacuum chamber with a pressure of $\sim$ 10$^{-6}$\,mbar. The end temperature is 16\,mK. Note however that the thermometer for this measurement is calibrated down to 22\,mK only. Temperature below this limit is obtained by extrapolating the calibration. \label{MC}}
\end{figure}

In order to minimize the heat flow into the sample from the bath, we used a feedback control of the bath temperature ($T_b$) to follow the sample temperature ($T_s$). Because the heat flow is $\dot{Q}=\kappa\,\Delta T$ where $\Delta T=T_b-T_s$ and $\kappa$ is the thermal conductance between the bath and the pellet, minimizing $\Delta T$ suppresses $\dot{Q}$ significantly. With the bath temperature of 2\,K, the pellet was slowly cooled to 2\,K at $\mu_0H=5$\,T. Then we swept magnetic field to zero with the 0.25\,T/min rate while giving the feedback control to the bath. As shown in Figure~\ref{MC}, the pellet is quickly cooled down to less than 22\,mK, which is the lowest limit of the thermometer calibration. By extrapolating this calibration to lower temperatures, we find the end temperature of about 16\,mK. A small spike appearing around $T=80$\,mK ($\mu_0 H=0.1$\,T) is probably due to magnetic flux pinning of the superconducting magnet. After reaching the lowest temperature, the sample warms up slowly due to the heat flow from the surroundings (the bath and radiation). Note that $T_b$ deviates from $T_s$ below 0.4\,K. This is because the cooling rate of $T_s$ is too fast for the bath to follow. The large temperature difference indicates a weak thermal conductance between the sample and the bath. Because of the difference there is finite heat flow into the sample. Therefore, under ideal adiabatic conditions the end temperature would be even lower than the extrapolated 16\,mK.

We note that there is no visible anomaly in sample temperatures other than the spike at 80\,mK. A phase transition of second (or first) order would cause a kink (or a local plateau). The absence of such an anomaly confirms the absence of magnetic ordering in \KBa{} at least down to 22\,mK and most likely down to about 16\,mK, the temperature several times lower than $|\theta|=60$\,mK. This behavior is in line with the geometrical frustration expected in \KBa{}. Assuming $\theta=-\frac32\mathcal J$ in a nearest-neighbor triangular antiferromagnet and $T_N/\mathcal J\simeq 0.2$ reported for Co-based frustrated materials~\cite{li2020}, we estimate $\mathcal J\simeq 40$\,mK and $T_N\simeq 8$\,mK, which would be comparable with the end temperature of 16\,mK achieved in our cooling experiment.

\begin{figure}
\includegraphics[width=\linewidth,keepaspectratio]{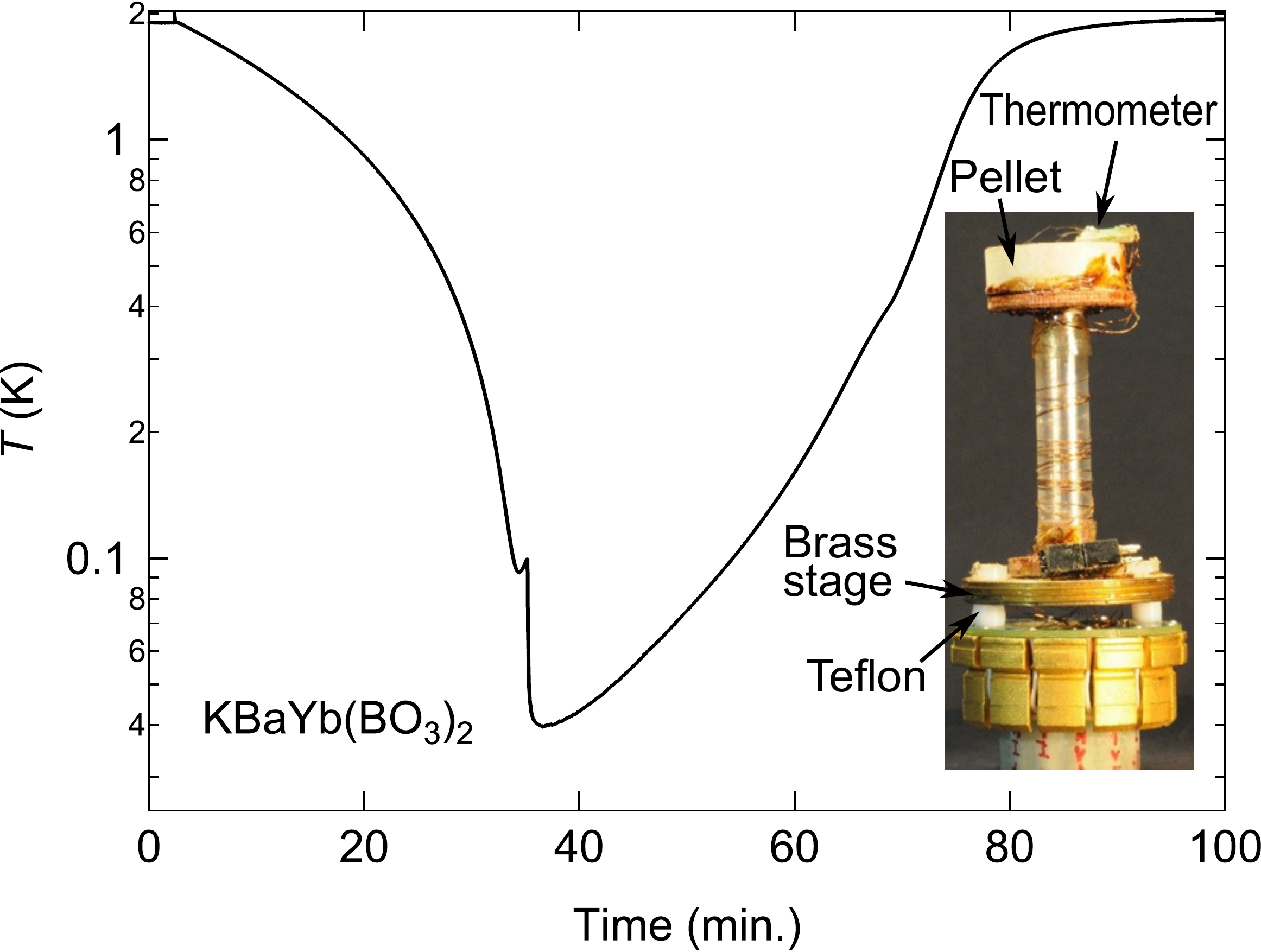}%
\caption{Cooling test with PPMS. The same sample as in Figure~\ref{MC} has been used. It is slowly cooled to $T$=2\,K at $\mu_{\rm 0}H$=5\,T through the weak thermal link to the PPMS puck kept at 2\,K. Magnetic field is swept from 5\,T to 0\,T with a rate of 0.15\,T/min. The inset shows a photo of the setup with a two-step thermal insulation from the puck (bath). The brass stage is fixed to the puck with two plastic screws having Teflon spacers. The sample pellet is further insulated from the brass stage by inserting a plastic straw. To reduce thermal conductance, 50\,$\mu$m-diameter 30\,cm-long resistive wires (manganin) are used for temperature measurement by the same resistive thermometer as in the test shown in Figure~\ref{MC}. This setup is shielded from radiation by a standard metallic cap of PPMS. \label{PPMS}}
\end{figure}

An immediate practical application of an ADR material would be its integration into commercial PPMS to achieve temperatures below 1.8\,K without using $^3$He. To this end, we constructed a stage where the sample pellet is placed on a tall plastic straw and thermally isolated from the heat bath (Figure~\ref{PPMS}). A resistive RuO$_2$ thermometer is glued on the pellet and connected with thin resistive manganin wires  to suppress heat flow. Then the setup is shielded by a metallic cap to reduce thermal radiation from the surroundings. Using the "high-vacuum mode" of the PPMS (pressure below $10^{-4}$\,mbar), we set the bath temperature to 2\,K at a magnetic field of 5\,T. Then the pellet temperature slowly approaches 2\,K by a weak thermal link through the manganin wires and plastic straw. After reaching 2\,K, the field is swept to zero with a rate of 0.15\,T/min. The pellet reaches temperatures below 40\,mK, with a spike in the sample temperature around 0.1\,K due to flux pinning of the superconducting magnet. The pellet warms up back to 2\,K in about 50 minutes by the weak heat conduction. This warming rate is slow enough for data collection of transport properties, such as electrical or Hall resistivity. Although the difference in setup has to be taken into account, the performance of our stage is better than the commercial ADR option for PPMS, which utilizes the paramagnetic salt, CPA, and reaches 100\,mK only~\cite{ppms}.

\section{Discussion}

We conclude that \KBa{} can be highly efficient in ADR while containing no water molecules and showing excellent stability upon both exposure to air and heating. Key parameters of the ADR materials are compared in Table~\ref{tab:comparison}. Figure~\ref{HC_and_Entropy} illustrates that even with the moderate field of 5\,T almost all entropy $R\ln 2$ of the lowest Kramers doublet is used for refrigeration. The same is true for conventional paramagnetic salts~\cite{Pobell-92}. Therefore, we list full entropies of the ground state as $S_{\rm GS}$. From the absolute values of $S_{\rm GS}$, \KBa{} is clearly competitive with MAS and FAA as "high-temperature" paramagnetic salts that develop magnetic order below $T_m$ of 170 and 30\,mK, respectively, and thus cannot cool below these temperatures. The lower-temperature ADR comes at the cost of the higher dilution of the magnetic ions and, therefore, lower $S_{\rm GS}$ in CPA and especially CMN. It is in fact natural that high density of the magnetic moments and low magnetic ordering temperature are unlikely to coexist. Nevertheless, \KBa{} combines these mutually exclusive properties by virtue of its magnetic frustration and structural randomness that both help to suppress $T_m$ well below the energy scale of magnetic interactions given by $|\theta|=60$\,mK. The material also shows an outstanding density of magnetic ions, 6.7~nm$^{-3}$, which is much higher than in any of the paramagnetic salts and would allow the design of more compact ADR apparatus. Even higher density has been reported for YbPt$_2$Sn and Yb$_3$Ga$_5$O$_{12}$ that, however, develop magnetic short-range order below 0.25\,K and long-range order below 54\,mK, respectively, which limits their end temperatures~\cite{Jang-NC15,YbGG}.

In summary, we demonstrated the H$_2$O-free refrigerant KBaYb(BO$_3$)$_2$ with the high entropy/volume ratio and excellent ADR performance of achieving at least 22\,mK and possibly temperatures below 16\,mK upon demagnetization from $\mu_0H=5$\,T at 2\,K. The absence of water molecules in this new refrigerant brings several additional advantages over conventional and commercially used paramagnetic salts. Delicate treatment of these materials is required to avoid degradation and ensure a good thermal contact by incorporating wires to the refrigerant pill, which makes production very advanced~\cite{BARTLETT2010582}. With KBaYb(BO$_3$)$_2$, improving thermal contact is much easier because rough treatments are possible. In the present study, for example, we ground the sample into fine powder and sintered the pressed pellet of a mixture with silver powder at high temperatures. Because of their low stability, water-containing paramagnetic salts have to be sealed air-tight and cannot be heated. The proposed material \KBa{} is free from all these problems. The chemical stability combined with the excellent cooling performance render \KBa{} ideal for ADR applications.

\section{Methods}
Powder samples of \KBa{} were prepared by a solid-state reaction of K$_2$CO$_3$, BaCO$_3$, Yb$_2$O$_3$, and H$_3$BO$_3$ taken in stoichiometric amounts, with the 6\% excess of H$_3$BO$_3$ and 2\% excess of K$_2$CO$_3$ and BaCO$_3$ due to their marginal loss upon heating. The reactants were carefully ground, placed into alumina crucibles, and kept at 200\,$^{\circ}$C for 6 hours to remove absorbed water. They were subsequently annealed at 700\,$^{\circ}$C for 24 hours, re-ground, and re-annealed at 900\,$^{\circ}$C for another 24 hours. All annealing steps were performed in air. We confirmed the desired crystal structure by X-ray diffraction with a small amount of 0.68\,wt.\% of the Yb$_2$O$_3$ impurity phase (See Supplementary Information). Sintered powders were ground and pressed into small pellets with the mass of 18.4 and 2.69\,mg used for measurements of magnetic susceptibility and specific heat, respectively.

Magnetic susceptibility was measured by an MPMS3 SQUID magnetometer from  Quantum Design equipped with a $^3$He-refrigerator. Measurements of specific heat in the temperature range from 12\,K down to 400\,mK were performed by using a Quantum Design, Physical Property Measurement System (PPMS), equipped with a $^3$He-refrigerator insert. Specific heat measurements of quasi-adiabatic heat pulse method at the lowest temperature down to 50\,mK and the tests for ADR with this material have been performed, using a $^3$He-$^4$He dilution refrigerator and PPMS. These tests are described in more detail in Section 2.

\medskip

%

%

\section*{Acknowledgements} 
We thank Andreas Honecker for fruitful discussions and organizing a very useful and informative workshop on “New perspectives for low-temperature refrigeration with advanced magnetocaloric materials” in 2018. The work in Augsburg was supported by the German Research Foundation (DFG) via the Project No.107745057 (TRR80) and by the Federal Ministry for Education and Research through the Sofja  Kovalevkaya Award of Alexander von Humboldt Foundation (AAT). The work in Tokai, Japan was supported by JSPS KAKENHI Grant Numbers,	20K22338 and 20KK0061.

\section*{Author  contributions}
Y.T., A.A.T. and P.G. conceived and designed the study. Y.T. and S.B. performed the specific heat measurements, the tests using dilution refrigerator and PPMS. K.K. and A.J. performed the magnetic susceptibility measurements. K.K. and A.J. prepared and characterized the samples. Y.T., S.B., A.A.T. and P.G. discussed the results. Y.T. and A.A.T. prepared the manuscript.

\section*{Competing interests}
A utility model which covers the fundamental aspects of the publication has been filed in Germany by the University of Augsburg. It is registered under the file number 20 2020 002 079. Inventors are Philipp Gegenwart, Alexander A. Tsirlin and Sebastian Bachus. All other authors declare no competing interests.

\newpage

\begin{table}
\caption{\label{tab:comparison}
Parameters of magnetic refrigeration materials: $T_{\rm m}$ is the critical temperature of magnetic order, $S_{\rm GS}$ is the entropy of the ground-state multiplet, and $R$ is the gas constant. The abbreviations stand for \\
MAS: Mn(NH$_4$)$_2$(SO$_4$)$_2\cdot$6H$_2$O (manganese ammonium sulfate~\cite{faa}) \\
FAA: NH$_4$Fe(SO$_4$)$_2\cdot$12H$_2$O\ (ferric ammonium alum~\cite{faa}) \\
CPA: KCr(SO$_4$)$_2\cdot$12H$_2$O\ (chromium potassium alum~\cite{cpa}) \\
CMN: Mg$_3$Ce$_2$(NO$_3$)$_{12}\cdot$24H$_2$O\ (cerium magnesium nitrate~\cite{cmn})
}
\begin{tabular}{l c c c c}
\hline\hline\\
{Material} & {$T_{\rm m}$} & {(mag. ion)/Vol.} & {$S_{\rm GS}$} & {$S_{\rm GS}$/vol.} \\
& {mK} & {nm$^{-3}$} && {mJ/K~cm$^3$} \\

\hline\\

 MAS~\cite{faa} & 170 & 2.8 & $R\ln6$ & 70 \\
 FAA~\cite{faa}  &  30 &  2.1 & $R\ln6$  &  53 \\
 CPA~\cite{cpa}) & 10 & 2.2 & $R\ln4$ & 42 \\
 CMN~\cite{cmn}) & 2 & 1.7 & $R\ln2$ & 16 \\
 \KBa{} & $<$22 & 6.7 & $R\ln2$ & 64 \\
YbPt$_2$Sn~\cite{Jang-NC15} & 250 & 12.9 & $R\ln2$ & 124 \\
Yb$_3$Ga$_5$O$_{12}$~\cite{YbGG} & 54 & 13.2 & $R\ln2$ & 124 \\

\hline
\end{tabular}
\end{table}

\widetext
\clearpage
\begin{center}

\textbf{\large Supplementary Information: Frustrated magnet for adiabatic demagnetization cooling to milli-Kelvin temperatures}

\end{center}

\setcounter{equation}{0}
\setcounter{figure}{0}
\setcounter{table}{0}
\setcounter{page}{1}
\setcounter{section}{0}
\makeatletter
\renewcommand{\theequation}{S\arabic{equation}}
\renewcommand{\thefigure}{S\arabic{figure}}

\section{Sample characterization}

\begin{figure}[h]
\begin{center}
\includegraphics[width=\textwidth]{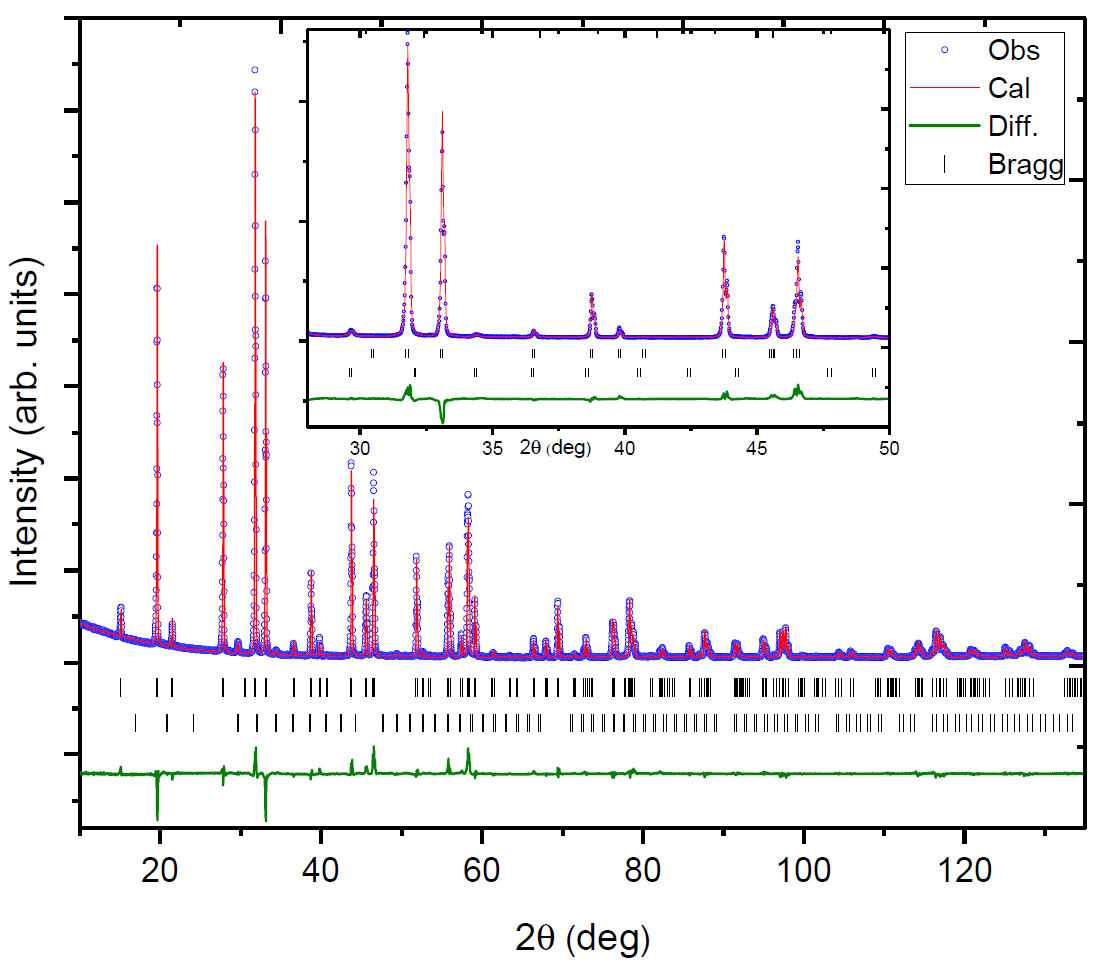}
\end{center}
\caption{\label{powder}
Rietveld refinement of X-ray diffraction data for \KBa{} at room temperature. Blue crosses are the data, red line is the calculated pattern, green line is the difference, and black vertical lines are the calculated Bragg reflections for \KBa{} (upper) and Yb$_2$O$_3$ (lower). The inset shows a blow-up in the angle range where several weak reflections from Yb$_2$O$_3$ are visible.
}
\end{figure}

X-ray powder diffraction pattern was recorded at room temperature using the Empyrean diffractometer from PANalytical (CuK$_{\alpha}$ radiation, reflection mode). Rietveld refinement (Fig.~\ref{powder}) confirmed the formation of \KBa{} with the lattice parameters $a=5.41120(2)$\,\r A and $c=17.5926(1)$\,\r A in good agreement with the earlier report~\cite{sanders2017}. Several very weak peaks of Yb$_2$O$_3$ were detected, corresponding to 0.75(1)\,wt.\% of the impurity phase.

\section{Magnetic Susceptibility}
\begin{figure}[h]
\begin{center}
\includegraphics[width=\textwidth]{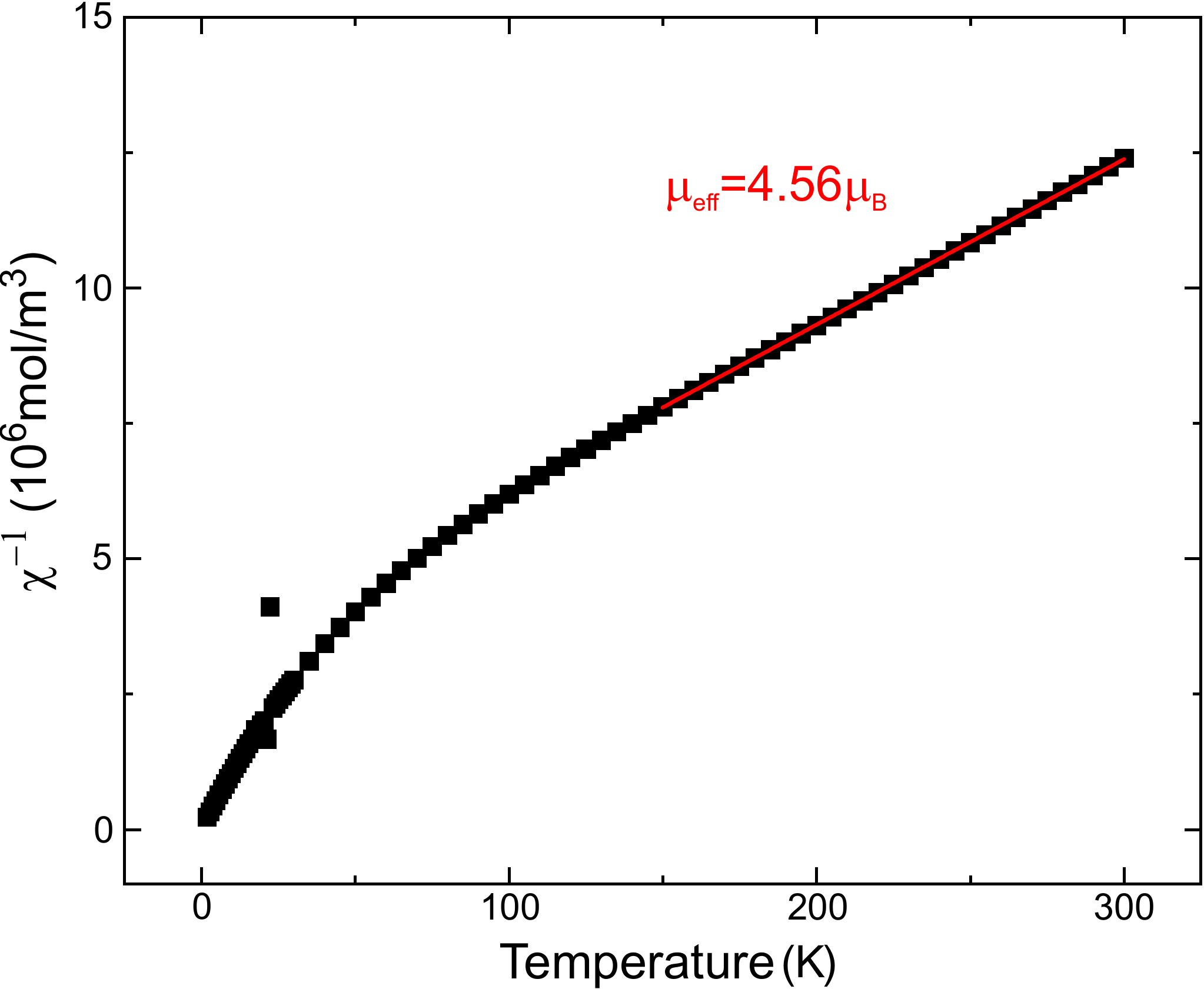}
\end{center}
\caption{\label{sus}
Inverse magnetic susceptibility of \KBa{}, plotted against temperature. Red line is a linear fit to the data in the temperature range from 150 to 300\,K, resulting in $\mu_{\rm eff}$=4.56\,$\mu_{\rm B}$ and Weiss temperature, $\theta=-105$\,K.
}

\end{figure}

Inverse magnetic susceptibility of \KBa{}, measured at 1\,T, as a function of temperature is shown in Fig.~\ref{sus}. The obtained effective magnetic moment from fitting the data at high temperatures, $\mu_{\rm eff}$=4.56$\mu_{\rm B}$, is in good agreement with 4.54\,$\mu_{\rm B}$ for free Yb$^{3+}$ ion.

\section{Contribution of the impurity phase}

\begin{figure}[h]
\begin{center}
\includegraphics[width=0.8\textwidth]{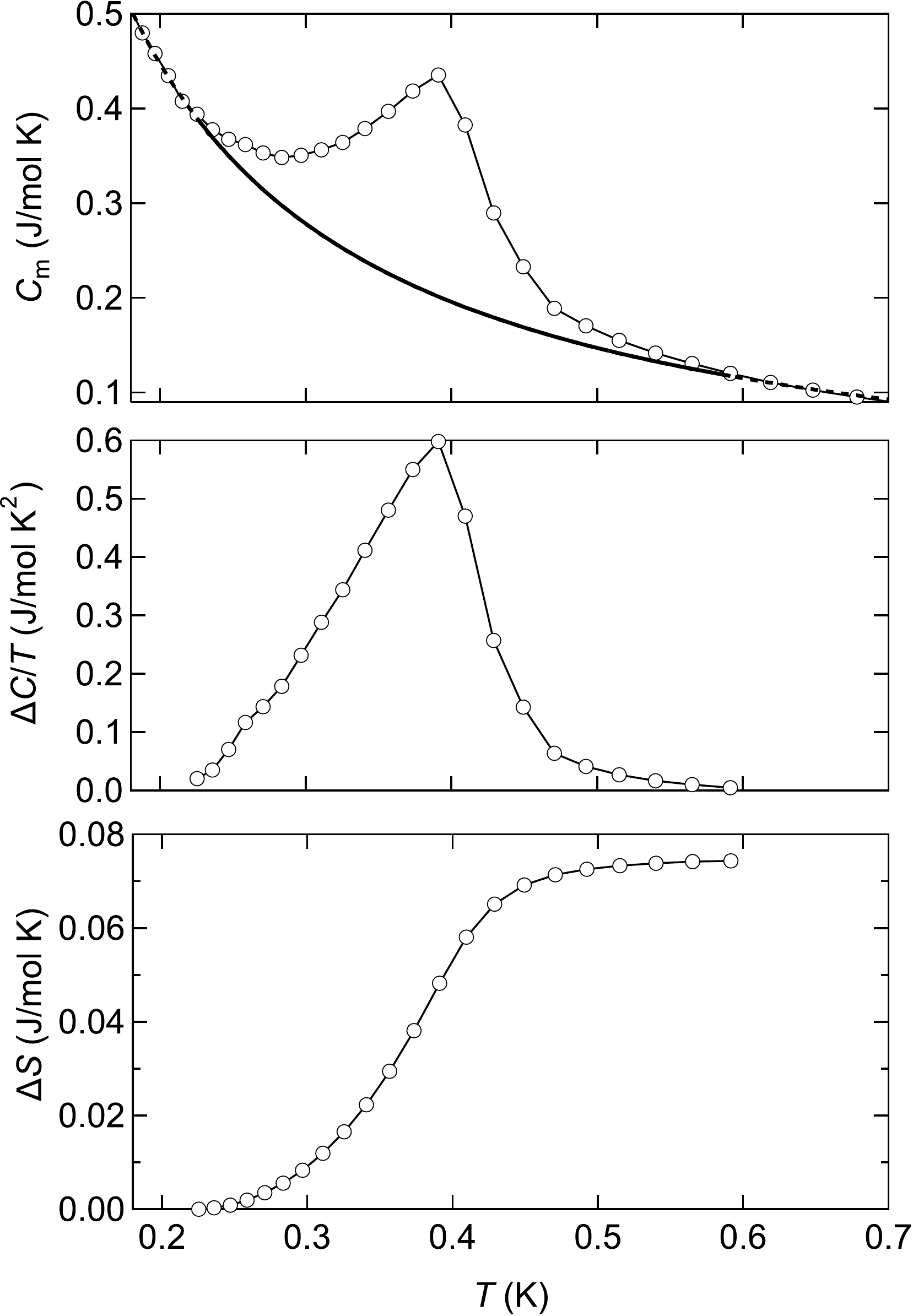}
\end{center}
\caption{\label{imp}
(a) Experimental magnetic specific heat ($C_{\rm m}$, open symbols) and interpolated specific heat of the main phase, \KBa{} (grey line). The difference is the impurity contribution. (b) The difference divided by temperature, $\Delta C$/$T$. (c) Impurity contribution to the entropy, obtained by integrating $\Delta C$/$T$.
}
\end{figure}

A weak anomaly is visible in our zero-field specific heat data around 0.4\,K. Here, we estimate magnetic entropy associated with this anomaly. We assume a smooth curve that connects the data points below and above the anomaly (see the gray line in Fig.~\ref{imp}a). The difference between the experimental data and this curve shows a $\lambda$-type anomaly in $\Delta C/T$ typical of a second-order phase transition (Fig.~\ref{imp}b). The corresponding entropy increment $\Delta S$ is obtained by integrating $\Delta C/T$ and amounts to 0.074\,J\,mol$^{-1}$\,K$^{-1}$, which is 1.3\,\% of $R\ln2$.

Considering that only 0.75(1)\,wt.\% of the Yb$_2$O$_3$ impurity is contained in the sample, and the magnetic ordering transition of Yb$_2$O$_3$ is at 2.3\,K~\cite{Yb2O3}, we infer that the anomaly should have a different origin. On the other hand, the small amount of entropy suggests that it is unrelated to \KBa{} itself. The most plausible explanation would be the presence of an amorphous or weakly crystalline impurity that is not seen in x-ray diffraction. Note also that a similar anomaly has been reported for the isostructural compound NaBaYb(BO$_3)_2$~\cite{guo2019b}.

\section{Low temperature extrapolation of thermometer calibration}
\begin{figure}[h]
\begin{center}
\includegraphics[width=\textwidth]{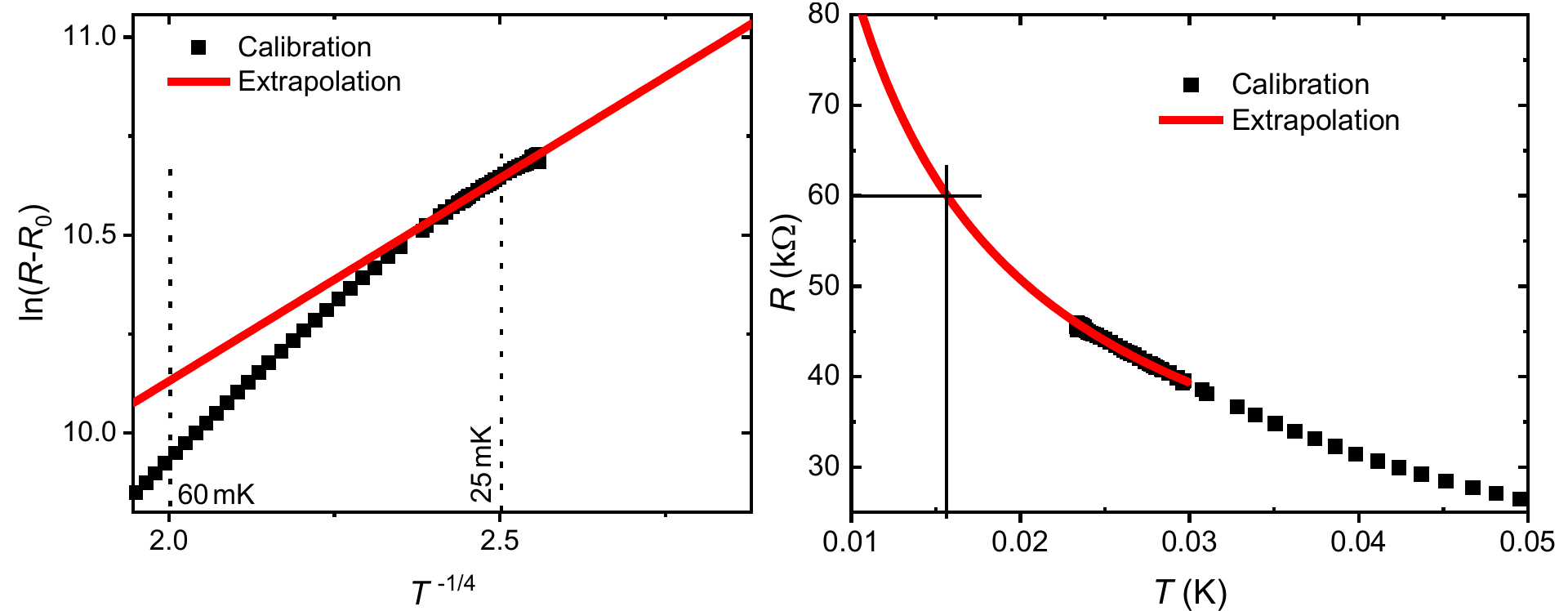}
\end{center}
\caption{\label{extrapo}
Resistive thermometer calibration and extrapolation to lower temperatures. The thermometer is Ruthenium oxide semiconductor. (left) $\ln(R-R_0)$ plotted against $T^{\rm -1/4}$ where $R_0$=2000\,$\Omega$ is the resistivity at room temperature. A linear in $T^{\rm -1/4}$ behavior of $\ln(R-R_0)$ is expected for variable-range hopping in doped semiconductors. Red line is a linear fit to the data. (right) Resistivity of the thermometer plotted against temperature in linear scale. Red line is an extrapolation obtained by the fitting. The resistance value of 60\,k$\Omega$ at the end temperature of the cooling cycle corresponds to 16\,mK, as shown with black lines.
}
\end{figure}

A ruthenium oxide thermometer is used for the tests of adiabatic demagnetization refrigeration with \KBa{}. It is calibrated down to 22\,mK. For temperatures below this lower limit, an extrapolation has to be used, as shown in Figure~\ref{extrapo}. Since $\ln(R-R_0)\sim T^{-1/4}$, with the resistance at room temperature $R_0$, is expected for variable-range hopping in semiconductors, we use a linear fit in Fig.~4(left) for the extrapolation. This corresponds to the final temperature not exceeding 16\,mK.

\newpage

\begin{table}[!ht]
 \caption{Atomic positions and atomic displacement parameters ($U_{\rm iso}$, in \r A$^2$) for KBaYb(BO$_3$)$_2$, as obtained from the Rietveld refinement. }
\begin{tabular}{c c c c c c}
\hline\
{Atoms} & {$x/a$} &{$y/b$} &{$z/c$} &{Occupancy} & {$U_{\rm iso}$} \\\hline
 Yb & 0 & 0 & 0 & 1 &  0.007(2)\\
 K & 0 & 0 & 0.7868(5) & 0.5 & 0.010(1)\\
 Ba & 0 & 0 & 0.7868(5) & 0.5 & 0.010(1)\\
 B & $\frac23$ & $\frac13$ & 0.7419(5) & 1 & 0.021(3)\\
 O & 0.5189(2) & 0.4811(2) & 0.7473(1) & 1 & 0.011(1)\\
 \hline
\end{tabular}

\end{table}

\begin{thebibliography}{37}%
\makeatletter
\providecommand \@ifxundefined [1]{%
 \@ifx{#1\undefined}
}%
\providecommand \@ifnum [1]{%
 \ifnum #1\expandafter \@firstoftwo
 \else \expandafter \@secondoftwo
 \fi
}%
\providecommand \@ifx [1]{%
 \ifx #1\expandafter \@firstoftwo
 \else \expandafter \@secondoftwo
 \fi
}%
\providecommand \natexlab [1]{#1}%
\providecommand \enquote  [1]{``#1''}%
\providecommand \bibnamefont  [1]{#1}%
\providecommand \bibfnamefont [1]{#1}%
\providecommand \citenamefont [1]{#1}%
\providecommand \href@noop [0]{\@secondoftwo}%
\providecommand \href [0]{\begingroup \@sanitize@url \@href}%
\providecommand \@href[1]{\@@startlink{#1}\@@href}%
\providecommand \@@href[1]{\endgroup#1\@@endlink}%
\providecommand \@sanitize@url [0]{\catcode `\\12\catcode `\$12\catcode
  `\&12\catcode `\#12\catcode `\^12\catcode `\_12\catcode `\%12\relax}%
\providecommand \@@startlink[1]{}%
\providecommand \@@endlink[0]{}%
\providecommand \url  [0]{\begingroup\@sanitize@url \@url }%
\providecommand \@url [1]{\endgroup\@href {#1}{\urlprefix }}%
\providecommand \urlprefix  [0]{URL }%
\providecommand \Eprint [0]{\href }%
\providecommand \doibase [0]{http://dx.doi.org/}%
\providecommand \selectlanguage [0]{\@gobble}%
\providecommand \bibinfo  [0]{\@secondoftwo}%
\providecommand \bibfield  [0]{\@secondoftwo}%
\providecommand \translation [1]{[#1]}%
\providecommand \BibitemOpen [0]{}%
\providecommand \bibitemStop [0]{}%
\providecommand \bibitemNoStop [0]{.\EOS\space}%
\providecommand \EOS [0]{\spacefactor3000\relax}%
\providecommand \BibitemShut  [1]{\csname bibitem#1\endcsname}%
\let\auto@bib@innerbib\@empty
\bibitem [{\citenamefont {Goerbig}(2011)}]{RevModPhys.83.1193}%
  \BibitemOpen
  \bibfield  {author} {\bibinfo {author} {\bibfnamefont {M.~O.}\ \bibnamefont
  {Goerbig}},\ }\bibfield  {title} {\enquote {\bibinfo {title} {Electronic
  properties of graphene in a strong magnetic field},}\ }\href {\doibase
  10.1103/RevModPhys.83.1193} {\bibfield  {journal} {\bibinfo  {journal} {Rev.
  Mod. Phys.}\ }\textbf {\bibinfo {volume} {83}},\ \bibinfo {pages}
  {1193--1243} (\bibinfo {year} {2011})}\BibitemShut {NoStop}%
\bibitem [{\citenamefont {Kamerlingh~Onnes}(1911)}]{SC}%
  \BibitemOpen
  \bibfield  {author} {\bibinfo {author} {\bibfnamefont {H.}~\bibnamefont
  {Kamerlingh~Onnes}},\ }\bibfield  {title} {\enquote {\bibinfo {title} {The
  discovery of superconductivity},}\ }\href@noop {} {\bibfield  {journal}
  {\bibinfo  {journal} {Commun. Phys. Lab. Univ. Leiden. Suppl.}\ }\textbf
  {\bibinfo {volume} {29}} (\bibinfo {year} {1911})}\BibitemShut {NoStop}%
\bibitem [{\citenamefont {Zhu}\ \emph {et~al.}(2011)\citenamefont {Zhu},
  \citenamefont {Saito}, \citenamefont {Kemp}, \citenamefont {Kakuyanagi},
  \citenamefont {Karimoto}, \citenamefont {Nakano}, \citenamefont {Munro},
  \citenamefont {Tokura}, \citenamefont {Everitt}, \citenamefont {Nemoto},
  \citenamefont {Kasu}, \citenamefont {Mizuochi},\ and\ \citenamefont
  {Semba}}]{Zhu2011}%
  \BibitemOpen
  \bibfield  {author} {\bibinfo {author} {\bibfnamefont {Xiaobo}\ \bibnamefont
  {Zhu}}, \bibinfo {author} {\bibfnamefont {Shiro}\ \bibnamefont {Saito}},
  \bibinfo {author} {\bibfnamefont {Alexander}\ \bibnamefont {Kemp}}, \bibinfo
  {author} {\bibfnamefont {Kosuke}\ \bibnamefont {Kakuyanagi}}, \bibinfo
  {author} {\bibfnamefont {Shin-ichi}\ \bibnamefont {Karimoto}}, \bibinfo
  {author} {\bibfnamefont {Hayato}\ \bibnamefont {Nakano}}, \bibinfo {author}
  {\bibfnamefont {William~J.}\ \bibnamefont {Munro}}, \bibinfo {author}
  {\bibfnamefont {Yasuhiro}\ \bibnamefont {Tokura}}, \bibinfo {author}
  {\bibfnamefont {Mark~S.}\ \bibnamefont {Everitt}}, \bibinfo {author}
  {\bibfnamefont {Kae}\ \bibnamefont {Nemoto}}, \bibinfo {author}
  {\bibfnamefont {Makoto}\ \bibnamefont {Kasu}}, \bibinfo {author}
  {\bibfnamefont {Norikazu}\ \bibnamefont {Mizuochi}}, \ and\ \bibinfo {author}
  {\bibfnamefont {Kouichi}\ \bibnamefont {Semba}},\ }\bibfield  {title}
  {\enquote {\bibinfo {title} {Coherent coupling of a superconducting flux
  qubit to an electron spin ensemble in diamond},}\ }\href {\doibase
  10.1038/nature10462} {\bibfield  {journal} {\bibinfo  {journal} {Nature}\
  }\textbf {\bibinfo {volume} {478}},\ \bibinfo {pages} {221--224} (\bibinfo
  {year} {2011})}\BibitemShut {NoStop}%
\bibitem [{\citenamefont {Irwin}(1995)}]{LTD}%
  \BibitemOpen
  \bibfield  {author} {\bibinfo {author} {\bibfnamefont {K.~D.}\ \bibnamefont
  {Irwin}},\ }\bibfield  {title} {\enquote {\bibinfo {title} {An application of
  electrothermal feedback for high resolution cryogenic particle detection},}\
  }\href {\doibase 10.1063/1.113674} {\bibfield  {journal} {\bibinfo  {journal}
  {Appl. Phys. Lett.}\ }\textbf {\bibinfo {volume} {66}},\ \bibinfo {pages}
  {1998--2000} (\bibinfo {year} {1995})}\BibitemShut {NoStop}%
\bibitem [{\citenamefont {Pobell}(1992)}]{Pobell-92}%
  \BibitemOpen
  \bibfield  {author} {\bibinfo {author} {\bibfnamefont {F.}~\bibnamefont
  {Pobell}},\ }\href@noop {} {\emph {\bibinfo {title} {Matter and Methods at
  Low Temperatures}}}\ (\bibinfo  {publisher} {Springer},\ \bibinfo {year}
  {1992})\BibitemShut {NoStop}%
\bibitem [{\citenamefont {Debye}(1926)}]{debye}%
  \BibitemOpen
  \bibfield  {author} {\bibinfo {author} {\bibfnamefont {P.}~\bibnamefont
  {Debye}},\ }\bibfield  {title} {\enquote {\bibinfo {title} {Einige
  {Bemerkungen} zur {Magnetisierung} bei tiefer {Temperatur}},}\ }\href
  {\doibase 10.1002/andp.19263862517} {\bibfield  {journal} {\bibinfo
  {journal} {Ann. Phys.}\ }\textbf {\bibinfo {volume} {386}},\ \bibinfo {pages}
  {1154--1160} (\bibinfo {year} {1926})}\BibitemShut {NoStop}%
\bibitem [{\citenamefont {Giauque}(1927)}]{giauque}%
  \BibitemOpen
  \bibfield  {author} {\bibinfo {author} {\bibfnamefont {W.~F.}\ \bibnamefont
  {Giauque}},\ }\bibfield  {title} {\enquote {\bibinfo {title} {A thermodynamic
  treatment of certain magnetic effects. a proposed method of producing
  temperatures considerably below $1^{\circ}$ absolute},}\ }\href {\doibase
  10.1021/ja01407a003} {\bibfield  {journal} {\bibinfo  {journal} {J. Amer.
  Chem. Soc.}\ }\textbf {\bibinfo {volume} {49}},\ \bibinfo {pages}
  {1864--1870} (\bibinfo {year} {1927})}\BibitemShut {NoStop}%
\bibitem [{\citenamefont {Shea}\ and\ \citenamefont
  {Morgan}(2010)}]{Shea-CRS10}%
  \BibitemOpen
  \bibfield  {author} {\bibinfo {author} {\bibfnamefont {D.}~\bibnamefont
  {Shea}}\ and\ \bibinfo {author} {\bibfnamefont {D..}\ \bibnamefont
  {Morgan}},\ }\href@noop {} {\emph {\bibinfo {title} {The {H}elium-3 shortage:
  supply, demand, and options for congress}}},\ \bibinfo {type} {Tech. Rep.}\
  \bibinfo {number} {R41419}\ (\bibinfo  {institution} {Congressional Research
  Service},\ \bibinfo {year} {2010})\BibitemShut {NoStop}%
\bibitem [{\citenamefont {Kouzes}\ and\ \citenamefont {Ely}(2010)}]{Kouzes}%
  \BibitemOpen
  \bibfield  {author} {\bibinfo {author} {\bibfnamefont {R.~T.}\ \bibnamefont
  {Kouzes}}\ and\ \bibinfo {author} {\bibfnamefont {J.~H.}\ \bibnamefont
  {Ely}},\ }\href
  {http://www.pnl.gov/main/publications/external/technical_reports/PNNL-19360.pdf}
  {\emph {\bibinfo {title} {Status summary of $^3${H}e and neutron detection
  alternatives for homeland security}}},\ \bibinfo {type} {Tech. Rep.}\
  \bibinfo {number} {PNNL-19360}\ (\bibinfo  {institution} {Pacific Northwest
  National Laboratory},\ \bibinfo {year} {2010})\BibitemShut {NoStop}%
\bibitem [{\citenamefont {Cho}(2009)}]{Cho-Science2009}%
  \BibitemOpen
  \bibfield  {author} {\bibinfo {author} {\bibfnamefont {A.}~\bibnamefont
  {Cho}},\ }\bibfield  {title} {\enquote {\bibinfo {title} {Helium-3 shortage
  could put freeze on low-temperature research},}\ }\href {\doibase
  {10.1126/science.326_778}} {\bibfield  {journal} {\bibinfo  {journal}
  {Science}\ }\textbf {\bibinfo {volume} {326}},\ \bibinfo {pages} {778--779}
  (\bibinfo {year} {2009})}\BibitemShut {NoStop}%
\bibitem [{\citenamefont {Wolf}\ \emph {et~al.}(2011)\citenamefont {Wolf},
  \citenamefont {Tsui}, \citenamefont {Jaiswal-Nagar}, \citenamefont {Tutsch},
  \citenamefont {Honecker}, \citenamefont {Removi{\'c}-Langer}, \citenamefont
  {Hofmann}, \citenamefont {Prokofiev}, \citenamefont {Assmus}, \citenamefont
  {Donath},\ and\ \citenamefont {Lang}}]{Wolf-PNAS11}%
  \BibitemOpen
  \bibfield  {author} {\bibinfo {author} {\bibfnamefont {B.}~\bibnamefont
  {Wolf}}, \bibinfo {author} {\bibfnamefont {Y.}~\bibnamefont {Tsui}}, \bibinfo
  {author} {\bibfnamefont {D.}~\bibnamefont {Jaiswal-Nagar}}, \bibinfo {author}
  {\bibfnamefont {U.}~\bibnamefont {Tutsch}}, \bibinfo {author} {\bibfnamefont
  {A.}~\bibnamefont {Honecker}}, \bibinfo {author} {\bibfnamefont
  {K.}~\bibnamefont {Removi{\'c}-Langer}}, \bibinfo {author} {\bibfnamefont
  {G.}~\bibnamefont {Hofmann}}, \bibinfo {author} {\bibfnamefont
  {A.}~\bibnamefont {Prokofiev}}, \bibinfo {author} {\bibfnamefont
  {W.}~\bibnamefont {Assmus}}, \bibinfo {author} {\bibfnamefont
  {G.}~\bibnamefont {Donath}}, \ and\ \bibinfo {author} {\bibfnamefont
  {M.}~\bibnamefont {Lang}},\ }\bibfield  {title} {\enquote {\bibinfo {title}
  {Magnetocaloric effect and magnetic cooling near a field-induced
  quantum-critical point},}\ }\href {\doibase 10.1073/pnas.1017047108}
  {\bibfield  {journal} {\bibinfo  {journal} {Proc. Nat. Acad. Sci.}\ }\textbf
  {\bibinfo {volume} {108}},\ \bibinfo {pages} {6862--6866} (\bibinfo {year}
  {2011})}\BibitemShut {NoStop}%
\bibitem [{\citenamefont {Jang}\ \emph {et~al.}(2015)\citenamefont {Jang},
  \citenamefont {Gruner}, \citenamefont {Steppke}, \citenamefont {Mitsumoto},
  \citenamefont {Geibel},\ and\ \citenamefont {Brando}}]{Jang-NC15}%
  \BibitemOpen
  \bibfield  {author} {\bibinfo {author} {\bibfnamefont {D.}~\bibnamefont
  {Jang}}, \bibinfo {author} {\bibfnamefont {T.}~\bibnamefont {Gruner}},
  \bibinfo {author} {\bibfnamefont {A.}~\bibnamefont {Steppke}}, \bibinfo
  {author} {\bibfnamefont {K.}~\bibnamefont {Mitsumoto}}, \bibinfo {author}
  {\bibfnamefont {C.}~\bibnamefont {Geibel}}, \ and\ \bibinfo {author}
  {\bibfnamefont {M.}~\bibnamefont {Brando}},\ }\bibfield  {title} {\enquote
  {\bibinfo {title} {Large magnetocaloric effect and adiabatic demagnetization
  refrigeration with {Y}b{P}t$_2${S}n},}\ }\href {\doibase 10.1038/ncomms9680}
  {\bibfield  {journal} {\bibinfo  {journal} {Nat. Commun.}\ }\textbf {\bibinfo
  {volume} {6}},\ \bibinfo {pages} {8680} (\bibinfo {year} {2015})}\BibitemShut
  {NoStop}%
\bibitem [{\citenamefont {{Paixao Brasiliano}}\ \emph
  {et~al.}(2020)\citenamefont {{Paixao Brasiliano}}, \citenamefont {Duval},
  \citenamefont {Marin}, \citenamefont {Bichaud}, \citenamefont {Brison},
  \citenamefont {Zhitomirsky},\ and\ \citenamefont {Luchier}}]{YbGG}%
  \BibitemOpen
  \bibfield  {author} {\bibinfo {author} {\bibfnamefont {Diego~Augusto}\
  \bibnamefont {{Paixao Brasiliano}}}, \bibinfo {author} {\bibfnamefont
  {Jean-Marc}\ \bibnamefont {Duval}}, \bibinfo {author} {\bibfnamefont
  {Christophe}\ \bibnamefont {Marin}}, \bibinfo {author} {\bibfnamefont
  {Emmanuelle}\ \bibnamefont {Bichaud}}, \bibinfo {author} {\bibfnamefont
  {Jean-Pascal}\ \bibnamefont {Brison}}, \bibinfo {author} {\bibfnamefont
  {Mike}\ \bibnamefont {Zhitomirsky}}, \ and\ \bibinfo {author} {\bibfnamefont
  {Nicolas}\ \bibnamefont {Luchier}},\ }\bibfield  {title} {\enquote {\bibinfo
  {title} {{YbGG} material for adiabatic demagnetization in the
  100 m{K}–3 {K} range},}\ }\href {\doibase
  10.1016/j.cryogenics.2019.103002} {\bibfield  {journal} {\bibinfo  {journal}
  {Cryogenics}\ }\textbf {\bibinfo {volume} {105}},\ \bibinfo {pages} {103002}
  (\bibinfo {year} {2020})}\BibitemShut {NoStop}%
\bibitem [{\citenamefont {Tokiwa}\ \emph {et~al.}(2016)\citenamefont {Tokiwa},
  \citenamefont {Piening}, \citenamefont {Jeevan}, \citenamefont
  {Bud{\textquoteright}ko}, \citenamefont {Canfield},\ and\ \citenamefont
  {Gegenwart}}]{Tokiwae1600835}%
  \BibitemOpen
  \bibfield  {author} {\bibinfo {author} {\bibfnamefont {Yoshifumi}\
  \bibnamefont {Tokiwa}}, \bibinfo {author} {\bibfnamefont {Boy}\ \bibnamefont
  {Piening}}, \bibinfo {author} {\bibfnamefont {Hirale~S.}\ \bibnamefont
  {Jeevan}}, \bibinfo {author} {\bibfnamefont {Sergey~L.}\ \bibnamefont
  {Bud{\textquoteright}ko}}, \bibinfo {author} {\bibfnamefont {Paul~C.}\
  \bibnamefont {Canfield}}, \ and\ \bibinfo {author} {\bibfnamefont {Philipp}\
  \bibnamefont {Gegenwart}},\ }\bibfield  {title} {\enquote {\bibinfo {title}
  {Super-heavy electron material as metallic refrigerant for adiabatic
  demagnetization cooling},}\ }\href {\doibase 10.1126/sciadv.1600835}
  {\bibfield  {journal} {\bibinfo  {journal} {Sci. Adv.}\ }\textbf {\bibinfo
  {volume} {2}},\ \bibinfo {pages} {e1600835} (\bibinfo {year}
  {2016})}\BibitemShut {NoStop}%
\bibitem [{\citenamefont {Evangelisti}\ \emph {et~al.}(2011)\citenamefont
  {Evangelisti}, \citenamefont {Roubeau}, \citenamefont {Palacios},
  \citenamefont {Camón}, \citenamefont {Hooper}, \citenamefont {Brechin},\
  and\ \citenamefont {Alonso}}]{Evangelisti}%
  \BibitemOpen
  \bibfield  {author} {\bibinfo {author} {\bibfnamefont {Marco}\ \bibnamefont
  {Evangelisti}}, \bibinfo {author} {\bibfnamefont {Olivier}\ \bibnamefont
  {Roubeau}}, \bibinfo {author} {\bibfnamefont {Elias}\ \bibnamefont
  {Palacios}}, \bibinfo {author} {\bibfnamefont {Agustín}\ \bibnamefont
  {Camón}}, \bibinfo {author} {\bibfnamefont {Thomas~N.}\ \bibnamefont
  {Hooper}}, \bibinfo {author} {\bibfnamefont {Euan~K.}\ \bibnamefont
  {Brechin}}, \ and\ \bibinfo {author} {\bibfnamefont {Juan~J.}\ \bibnamefont
  {Alonso}},\ }\bibfield  {title} {\enquote {\bibinfo {title} {Cryogenic
  magnetocaloric effect in a ferromagnetic molecular dimer},}\ }\href {\doibase
  10.1002/anie.201102640} {\bibfield  {journal} {\bibinfo  {journal} {Angew.
  Chem. Int. Ed.}\ }\textbf {\bibinfo {volume} {50}},\ \bibinfo {pages}
  {6606--6609} (\bibinfo {year} {2011})}\BibitemShut {NoStop}%
\bibitem [{\citenamefont {Baniodeh}\ \emph {et~al.}(2018)\citenamefont
  {Baniodeh}, \citenamefont {Magnani}, \citenamefont {Lan}, \citenamefont
  {Buth}, \citenamefont {Anson}, \citenamefont {Richter}, \citenamefont
  {Affronte}, \citenamefont {Schnack},\ and\ \citenamefont
  {Powell}}]{Baniodeh}%
  \BibitemOpen
  \bibfield  {author} {\bibinfo {author} {\bibfnamefont {Amer}\ \bibnamefont
  {Baniodeh}}, \bibinfo {author} {\bibfnamefont {Nicola}\ \bibnamefont
  {Magnani}}, \bibinfo {author} {\bibfnamefont {Yanhua}\ \bibnamefont {Lan}},
  \bibinfo {author} {\bibfnamefont {Gernot}\ \bibnamefont {Buth}}, \bibinfo
  {author} {\bibfnamefont {Christopher~E.}\ \bibnamefont {Anson}}, \bibinfo
  {author} {\bibfnamefont {Johannes}\ \bibnamefont {Richter}}, \bibinfo
  {author} {\bibfnamefont {Marco}\ \bibnamefont {Affronte}}, \bibinfo {author}
  {\bibfnamefont {J{\"u}rgen}\ \bibnamefont {Schnack}}, \ and\ \bibinfo
  {author} {\bibfnamefont {Annie~K.}\ \bibnamefont {Powell}},\ }\bibfield
  {title} {\enquote {\bibinfo {title} {High spin cycles: topping the spin
  record for a single molecule verging on quantum criticality},}\ }\href
  {\doibase 10.1038/s41535-018-0082-7} {\bibfield  {journal} {\bibinfo
  {journal} {npj Quantum Mater.}\ }\textbf {\bibinfo {volume} {3}},\ \bibinfo
  {pages} {10} (\bibinfo {year} {2018})}\BibitemShut {NoStop}%
\bibitem [{\citenamefont {Zhitomirsky}(2003)}]{Zhitomirsky}%
  \BibitemOpen
  \bibfield  {author} {\bibinfo {author} {\bibfnamefont {M.~E.}\ \bibnamefont
  {Zhitomirsky}},\ }\bibfield  {title} {\enquote {\bibinfo {title} {Enhanced
  magnetocaloric effect in frustrated magnets},}\ }\href {\doibase
  10.1103/PhysRevB.67.104421} {\bibfield  {journal} {\bibinfo  {journal} {Phys.
  Rev. B}\ }\textbf {\bibinfo {volume} {67}},\ \bibinfo {pages} {104421}
  (\bibinfo {year} {2003})}\BibitemShut {NoStop}%
\bibitem [{\citenamefont {Wolf}\ \emph {et~al.}(2016)\citenamefont {Wolf},
  \citenamefont {Tutsch}, \citenamefont {Dörschug}, \citenamefont {Krellner},
  \citenamefont {Ritter}, \citenamefont {Assmus},\ and\ \citenamefont
  {Lang}}]{Er227}%
  \BibitemOpen
  \bibfield  {author} {\bibinfo {author} {\bibfnamefont {B.}~\bibnamefont
  {Wolf}}, \bibinfo {author} {\bibfnamefont {U.}~\bibnamefont {Tutsch}},
  \bibinfo {author} {\bibfnamefont {S.}~\bibnamefont {Dörschug}}, \bibinfo
  {author} {\bibfnamefont {C.}~\bibnamefont {Krellner}}, \bibinfo {author}
  {\bibfnamefont {F.}~\bibnamefont {Ritter}}, \bibinfo {author} {\bibfnamefont
  {W.}~\bibnamefont {Assmus}}, \ and\ \bibinfo {author} {\bibfnamefont
  {M.}~\bibnamefont {Lang}},\ }\bibfield  {title} {\enquote {\bibinfo {title}
  {Magnetic cooling close to a quantum phase transition—the case of
  {E}r$_2${T}i$_2${O}$_7$},}\ }\href {\doibase 10.1073/pnas.1017047108}
  {\bibfield  {journal} {\bibinfo  {journal} {Journal of Applied Physics}\
  }\textbf {\bibinfo {volume} {120}},\ \bibinfo {pages} {142112} (\bibinfo
  {year} {2016})}\BibitemShut {NoStop}%
\bibitem [{\citenamefont {Hu}\ and\ \citenamefont {Du}(2008)}]{Hu_2008}%
  \BibitemOpen
  \bibfield  {author} {\bibinfo {author} {\bibfnamefont {Yong}\ \bibnamefont
  {Hu}}\ and\ \bibinfo {author} {\bibfnamefont {An}~\bibnamefont {Du}},\
  }\bibfield  {title} {\enquote {\bibinfo {title} {Magnetization behavior and
  magnetic entropy change of frustrated ising antiferromagnets on two- and
  three-dimensional lattices},}\ }\href {\doibase
  10.1088/0953-8984/20/12/125225} {\bibfield  {journal} {\bibinfo  {journal}
  {Journal of Physics: Condensed Matter}\ }\textbf {\bibinfo {volume} {20}},\
  \bibinfo {pages} {125225} (\bibinfo {year} {2008})}\BibitemShut {NoStop}%
\bibitem [{\citenamefont {Shirron}\ \emph {et~al.}(2004)\citenamefont
  {Shirron}, \citenamefont {Canavan}, \citenamefont {DiPirro}, \citenamefont
  {Francis}, \citenamefont {Jackson}, \citenamefont {Tuttle}, \citenamefont
  {King},\ and\ \citenamefont {Grabowski}}]{SHIRRON2004581}%
  \BibitemOpen
  \bibfield  {author} {\bibinfo {author} {\bibfnamefont {Peter}\ \bibnamefont
  {Shirron}}, \bibinfo {author} {\bibfnamefont {Ed}~\bibnamefont {Canavan}},
  \bibinfo {author} {\bibfnamefont {Michael}\ \bibnamefont {DiPirro}}, \bibinfo
  {author} {\bibfnamefont {John}\ \bibnamefont {Francis}}, \bibinfo {author}
  {\bibfnamefont {Michael}\ \bibnamefont {Jackson}}, \bibinfo {author}
  {\bibfnamefont {James}\ \bibnamefont {Tuttle}}, \bibinfo {author}
  {\bibfnamefont {Todd}\ \bibnamefont {King}}, \ and\ \bibinfo {author}
  {\bibfnamefont {Matt}\ \bibnamefont {Grabowski}},\ }\bibfield  {title}
  {\enquote {\bibinfo {title} {Development of a cryogen-free continuous {ADR}
  for the {constellation-X} mission},}\ }\href {\doibase
  10.1016/j.cryogenics.2003.11.011} {\bibfield  {journal} {\bibinfo  {journal}
  {Cryogenics}\ }\textbf {\bibinfo {volume} {44}},\ \bibinfo {pages} {581--588}
  (\bibinfo {year} {2004})}\BibitemShut {NoStop}%
\bibitem [{\citenamefont {Bartlett}\ \emph {et~al.}(2010)\citenamefont
  {Bartlett}, \citenamefont {Hardy}, \citenamefont {Hepburn}, \citenamefont
  {Brockley-Blatt}, \citenamefont {Coker}, \citenamefont {Crofts},
  \citenamefont {Winter}, \citenamefont {Milward}, \citenamefont
  {Stafford-Allen}, \citenamefont {Brownhill}, \citenamefont {Reed},
  \citenamefont {Linder},\ and\ \citenamefont {Rando}}]{BARTLETT2010582}%
  \BibitemOpen
  \bibfield  {author} {\bibinfo {author} {\bibfnamefont {J.}~\bibnamefont
  {Bartlett}}, \bibinfo {author} {\bibfnamefont {G.}~\bibnamefont {Hardy}},
  \bibinfo {author} {\bibfnamefont {I.D.}\ \bibnamefont {Hepburn}}, \bibinfo
  {author} {\bibfnamefont {C.}~\bibnamefont {Brockley-Blatt}}, \bibinfo
  {author} {\bibfnamefont {P.}~\bibnamefont {Coker}}, \bibinfo {author}
  {\bibfnamefont {E.}~\bibnamefont {Crofts}}, \bibinfo {author} {\bibfnamefont
  {B.}~\bibnamefont {Winter}}, \bibinfo {author} {\bibfnamefont
  {S.}~\bibnamefont {Milward}}, \bibinfo {author} {\bibfnamefont
  {R.}~\bibnamefont {Stafford-Allen}}, \bibinfo {author} {\bibfnamefont
  {M.}~\bibnamefont {Brownhill}}, \bibinfo {author} {\bibfnamefont
  {J.}~\bibnamefont {Reed}}, \bibinfo {author} {\bibfnamefont {M.}~\bibnamefont
  {Linder}}, \ and\ \bibinfo {author} {\bibfnamefont {N.}~\bibnamefont
  {Rando}},\ }\bibfield  {title} {\enquote {\bibinfo {title} {Improved
  performance of an engineering model cryogen free double adiabatic
  demagnetization refrigerator},}\ }\href {\doibase
  10.1016/j.cryogenics.2010.02.024} {\bibfield  {journal} {\bibinfo  {journal}
  {Cryogenics}\ }\textbf {\bibinfo {volume} {50}},\ \bibinfo {pages} {582 --
  590} (\bibinfo {year} {2010})}\BibitemShut {NoStop}%
\bibitem [{\citenamefont {kiutra}()}]{kiutra}%
  \BibitemOpen
  \bibfield  {author} {\bibinfo {author} {\bibnamefont {kiutra}},\ }\href@noop
  {} {\enquote {\bibinfo {title} {Cryogen-free magnetic refrigeration},}\
  }\bibinfo {howpublished} {\url{https://kiutra.com/technology/}}\BibitemShut
  {NoStop}%
\bibitem [{\citenamefont {Daniels}\ \emph {et~al.}(1954)\citenamefont
  {Daniels}, \citenamefont {Kurti},\ and\ \citenamefont {Simon}}]{cpa}%
  \BibitemOpen
  \bibfield  {author} {\bibinfo {author} {\bibfnamefont {J.~M.}\ \bibnamefont
  {Daniels}}, \bibinfo {author} {\bibfnamefont {Nicholas}\ \bibnamefont
  {Kurti}}, \ and\ \bibinfo {author} {\bibfnamefont {Franz~Eugen}\ \bibnamefont
  {Simon}},\ }\bibfield  {title} {\enquote {\bibinfo {title} {The thermal and
  magnetic properties of chromium potassium alum below 0.1\,{K}},}\ }\href
  {\doibase 10.1098/rspa.1954.0018} {\bibfield  {journal} {\bibinfo  {journal}
  {Proc. Roy. Soc. Lond. A}\ }\textbf {\bibinfo {volume} {221}},\ \bibinfo
  {pages} {243--256} (\bibinfo {year} {1954})}\BibitemShut {NoStop}%
\bibitem [{\citenamefont {Vilches}\ and\ \citenamefont
  {Wheatley}(1966{\natexlab{a}})}]{Viches-PR66}%
  \BibitemOpen
  \bibfield  {author} {\bibinfo {author} {\bibfnamefont {O.~E.}\ \bibnamefont
  {Vilches}}\ and\ \bibinfo {author} {\bibfnamefont {J.~C.}\ \bibnamefont
  {Wheatley}},\ }\bibfield  {title} {\enquote {\bibinfo {title} {Measurements
  of the specific heats of three magnetic salts at low temperatures},}\ }\href
  {\doibase 10.1103/PhysRev.148.509} {\bibfield  {journal} {\bibinfo  {journal}
  {Phys. Rev.}\ }\textbf {\bibinfo {volume} {148}},\ \bibinfo {pages}
  {509--516} (\bibinfo {year} {1966}{\natexlab{a}})}\BibitemShut {NoStop}%
\bibitem [{\citenamefont {Li}\ \emph {et~al.}(2020)\citenamefont {Li},
  \citenamefont {Gegenwart},\ and\ \citenamefont {Tsirlin}}]{li2020}%
  \BibitemOpen
  \bibfield  {author} {\bibinfo {author} {\bibfnamefont {Y.}~\bibnamefont
  {Li}}, \bibinfo {author} {\bibfnamefont {P.}~\bibnamefont {Gegenwart}}, \
  and\ \bibinfo {author} {\bibfnamefont {A.~A.}\ \bibnamefont {Tsirlin}},\
  }\bibfield  {title} {\enquote {\bibinfo {title} {Spin liquids in
  geometrically perfect triangular antiferromagnets},}\ }\href {\doibase
  10.1088/1361-648X/ab724e} {\bibfield  {journal} {\bibinfo  {journal} {J.
  Phys.: Condens. Matter}\ }\textbf {\bibinfo {volume} {32}},\ \bibinfo {pages}
  {224004} (\bibinfo {year} {2020})}\BibitemShut {NoStop}%
\bibitem [{\citenamefont {Sanders}\ \emph {et~al.}(2017)\citenamefont
  {Sanders}, \citenamefont {Cevallos},\ and\ \citenamefont
  {Cava}}]{sanders2017}%
  \BibitemOpen
  \bibfield  {author} {\bibinfo {author} {\bibfnamefont {M.~B.}\ \bibnamefont
  {Sanders}}, \bibinfo {author} {\bibfnamefont {F.~A.}\ \bibnamefont
  {Cevallos}}, \ and\ \bibinfo {author} {\bibfnamefont {R.~J.}\ \bibnamefont
  {Cava}},\ }\bibfield  {title} {\enquote {\bibinfo {title} {Magnetism in the
  {KBaRE(BO$_3)_2$ (RE = Sm, Eu, Gd, Tb, Dy, Ho, Er, Tm, Yb, Lu)} series:
  materials with a triangular rare earth lattice},}\ }\href {\doibase
  10.1088/2053-1591/aa60a2} {\bibfield  {journal} {\bibinfo  {journal} {Mater.
  Res. Express}\ }\textbf {\bibinfo {volume} {4}},\ \bibinfo {pages} {036102}
  (\bibinfo {year} {2017})}\BibitemShut {NoStop}%
\bibitem [{\citenamefont {Guo}\ \emph {et~al.}(2019{\natexlab{a}})\citenamefont
  {Guo}, \citenamefont {Kong}, \citenamefont {{Alex Cevallos}}, \citenamefont
  {Stolze},\ and\ \citenamefont {Cava}}]{guo2019a}%
  \BibitemOpen
  \bibfield  {author} {\bibinfo {author} {\bibfnamefont {S.}~\bibnamefont
  {Guo}}, \bibinfo {author} {\bibfnamefont {T.}~\bibnamefont {Kong}}, \bibinfo
  {author} {\bibfnamefont {F.}~\bibnamefont {{Alex Cevallos}}}, \bibinfo
  {author} {\bibfnamefont {K.}~\bibnamefont {Stolze}}, \ and\ \bibinfo {author}
  {\bibfnamefont {R.~J.}\ \bibnamefont {Cava}},\ }\bibfield  {title} {\enquote
  {\bibinfo {title} {Crystal growth, crystal structure and anisotropic magnetic
  properties of {KBaR(BO$_3)_2$ (R = Y, Gd, Tb, Dy, Ho, Tm, Yb, and Lu)}
  triangular lattice materials},}\ }\href {\doibase 10.1016/j.jmmm.2018.10.037}
  {\bibfield  {journal} {\bibinfo  {journal} {J. Magn. Magn. Mater.}\ }\textbf
  {\bibinfo {volume} {472}},\ \bibinfo {pages} {104--110} (\bibinfo {year}
  {2019}{\natexlab{a}})}\BibitemShut {NoStop}%
\bibitem [{\citenamefont {Guo}\ \emph {et~al.}(2019{\natexlab{b}})\citenamefont
  {Guo}, \citenamefont {Ghasemi}, \citenamefont {Broholm},\ and\ \citenamefont
  {Cava}}]{guo2019b}%
  \BibitemOpen
  \bibfield  {author} {\bibinfo {author} {\bibfnamefont {S.}~\bibnamefont
  {Guo}}, \bibinfo {author} {\bibfnamefont {A.}~\bibnamefont {Ghasemi}},
  \bibinfo {author} {\bibfnamefont {C.~L.}\ \bibnamefont {Broholm}}, \ and\
  \bibinfo {author} {\bibfnamefont {R.~J.}\ \bibnamefont {Cava}},\ }\bibfield
  {title} {\enquote {\bibinfo {title} {Magnetism on ideal triangular lattices
  in {NaBaYb(BO$_3)_2$}},}\ }\href {\doibase 10.1103/PhysRevMaterials.3.094404}
  {\bibfield  {journal} {\bibinfo  {journal} {Phys. Rev. Materials}\ }\textbf
  {\bibinfo {volume} {3}},\ \bibinfo {pages} {094404} (\bibinfo {year}
  {2019}{\natexlab{b}})}\BibitemShut {NoStop}%
\bibitem [{\citenamefont {Guo}\ \emph {et~al.}(2019{\natexlab{c}})\citenamefont
  {Guo}, \citenamefont {Kong}, \citenamefont {Xie}, \citenamefont {Nguyen},
  \citenamefont {Stolze}, \citenamefont {{Alex Cevallos}},\ and\ \citenamefont
  {Cava}}]{guo2019c}%
  \BibitemOpen
  \bibfield  {author} {\bibinfo {author} {\bibfnamefont {S.}~\bibnamefont
  {Guo}}, \bibinfo {author} {\bibfnamefont {T.}~\bibnamefont {Kong}}, \bibinfo
  {author} {\bibfnamefont {W.}~\bibnamefont {Xie}}, \bibinfo {author}
  {\bibfnamefont {L.}~\bibnamefont {Nguyen}}, \bibinfo {author} {\bibfnamefont
  {K.}~\bibnamefont {Stolze}}, \bibinfo {author} {\bibfnamefont
  {F.}~\bibnamefont {{Alex Cevallos}}}, \ and\ \bibinfo {author} {\bibfnamefont
  {R.~J.}\ \bibnamefont {Cava}},\ }\bibfield  {title} {\enquote {\bibinfo
  {title} {Triangular rare-earth lattice materials {RbBaR(BO$_3)_2$ (R = Y,
  Gd-Yb)} and comparison to the {KBaR(BO$_3$)$_2$} analogs},}\ }\href {\doibase
  10.1021/acs.inorgchem.8b03372} {\bibfield  {journal} {\bibinfo  {journal}
  {Inorg. Chem.}\ }\textbf {\bibinfo {volume} {58}},\ \bibinfo {pages}
  {3308--3315} (\bibinfo {year} {2019}{\natexlab{c}})}\BibitemShut {NoStop}%
\bibitem [{\citenamefont {Guo}\ \emph {et~al.}(2019{\natexlab{d}})\citenamefont
  {Guo}, \citenamefont {Kong},\ and\ \citenamefont {Cava}}]{guo2019d}%
  \BibitemOpen
  \bibfield  {author} {\bibinfo {author} {\bibfnamefont {S.}~\bibnamefont
  {Guo}}, \bibinfo {author} {\bibfnamefont {T.}~\bibnamefont {Kong}}, \ and\
  \bibinfo {author} {\bibfnamefont {R.~J.}\ \bibnamefont {Cava}},\ }\bibfield
  {title} {\enquote {\bibinfo {title} {{NaBaR(BO$_3)_2$ (R = Dy, Ho, Er and
  Tm)}: structurally perfect triangular lattice materials with two rare earth
  layers},}\ }\href {\doibase 10.1088/2053-1591/ab3d8e} {\bibfield  {journal}
  {\bibinfo  {journal} {Mater. Res. Express}\ }\textbf {\bibinfo {volume}
  {6}},\ \bibinfo {pages} {106110} (\bibinfo {year}
  {2019}{\natexlab{d}})}\BibitemShut {NoStop}%
\bibitem [{\citenamefont {Sosin}\ \emph {et~al.}(2005)\citenamefont {Sosin},
  \citenamefont {Prozorova}, \citenamefont {Smirnov}, \citenamefont {Golov},
  \citenamefont {Berkutov}, \citenamefont {Petrenko}, \citenamefont
  {Balakrishnan},\ and\ \citenamefont {Zhitomirsky}}]{GTO}%
  \BibitemOpen
  \bibfield  {author} {\bibinfo {author} {\bibfnamefont {S.~S.}\ \bibnamefont
  {Sosin}}, \bibinfo {author} {\bibfnamefont {L.~A.}\ \bibnamefont
  {Prozorova}}, \bibinfo {author} {\bibfnamefont {A.~I.}\ \bibnamefont
  {Smirnov}}, \bibinfo {author} {\bibfnamefont {A.~I.}\ \bibnamefont {Golov}},
  \bibinfo {author} {\bibfnamefont {I.~B.}\ \bibnamefont {Berkutov}}, \bibinfo
  {author} {\bibfnamefont {O.~A.}\ \bibnamefont {Petrenko}}, \bibinfo {author}
  {\bibfnamefont {G.}~\bibnamefont {Balakrishnan}}, \ and\ \bibinfo {author}
  {\bibfnamefont {M.~E.}\ \bibnamefont {Zhitomirsky}},\ }\bibfield  {title}
  {\enquote {\bibinfo {title} {Magnetocaloric effect in pyrochlore
  antiferromagnet {Gd}$_2${Ti}$_2${O}$_7$},}\ }\href@noop {} {\bibfield
  {journal} {\bibinfo  {journal} {Phys. Rev. B}\ }\textbf {\bibinfo {volume}
  {71}},\ \bibinfo {pages} {094413} (\bibinfo {year} {2005})}\BibitemShut
  {NoStop}%
\bibitem [{\citenamefont {Aoki}\ and\ \citenamefont {Flouquet}(2014)}]{ppms}%
  \BibitemOpen
  \bibfield  {author} {\bibinfo {author} {\bibfnamefont {Dai}\ \bibnamefont
  {Aoki}}\ and\ \bibinfo {author} {\bibfnamefont {Jacques}\ \bibnamefont
  {Flouquet}},\ }\bibfield  {title} {\enquote {\bibinfo {title}
  {Superconductivity and ferromagnetic quantum criticality in uranium
  compounds},}\ }\href {\doibase 10.7566/JPSJ.83.061011} {\bibfield  {journal}
  {\bibinfo  {journal} {J. Phys. Soc. Jpn.}\ }\textbf {\bibinfo {volume}
  {83}},\ \bibinfo {pages} {061011} (\bibinfo {year} {2014})}\BibitemShut
  {NoStop}%
\bibitem [{\citenamefont {Balents}(2010)}]{balents-nature10}%
  \BibitemOpen
  \bibfield  {author} {\bibinfo {author} {\bibfnamefont {Leon}\ \bibnamefont
  {Balents}},\ }\bibfield  {title} {\enquote {\bibinfo {title} {Spin liquids in
  frustrated magnets},}\ }\href {\doibase 10.1038/nature08917} {\bibfield
  {journal} {\bibinfo  {journal} {Nature}\ }\textbf {\bibinfo {volume} {464}},\
  \bibinfo {pages} {199--208} (\bibinfo {year} {2010})}\BibitemShut {NoStop}%
\bibitem [{\citenamefont {Dey}\ \emph {et~al.}(2020)\citenamefont {Dey},
  \citenamefont {Andrade},\ and\ \citenamefont {Vojta}}]{dey2020}%
  \BibitemOpen
  \bibfield  {author} {\bibinfo {author} {\bibfnamefont {S.}~\bibnamefont
  {Dey}}, \bibinfo {author} {\bibfnamefont {E.~C.}\ \bibnamefont {Andrade}}, \
  and\ \bibinfo {author} {\bibfnamefont {M.}~\bibnamefont {Vojta}},\ }\bibfield
   {title} {\enquote {\bibinfo {title} {Destruction of long-range order in
  noncollinear two-dimensional antiferromagnets by random-bond disorder},}\
  }\href {\doibase 10.1103/PhysRevB.101.020411} {\bibfield  {journal} {\bibinfo
   {journal} {Phys. Rev. B}\ }\textbf {\bibinfo {volume} {101}},\ \bibinfo
  {pages} {020411(R)} (\bibinfo {year} {2020})}\BibitemShut {NoStop}%
\bibitem [{\citenamefont {Vilches}\ and\ \citenamefont
  {Wheatley}(1966{\natexlab{b}})}]{faa}%
  \BibitemOpen
  \bibfield  {author} {\bibinfo {author} {\bibfnamefont {O.~E.}\ \bibnamefont
  {Vilches}}\ and\ \bibinfo {author} {\bibfnamefont {J.~C.}\ \bibnamefont
  {Wheatley}},\ }\bibfield  {title} {\enquote {\bibinfo {title} {Measurements
  of the specific heats of three magnetic salts at low temperatures},}\ }\href
  {\doibase 10.1103/PhysRev.148.509} {\bibfield  {journal} {\bibinfo  {journal}
  {Phys. Rev.}\ }\textbf {\bibinfo {volume} {148}},\ \bibinfo {pages}
  {509--516} (\bibinfo {year} {1966}{\natexlab{b}})}\BibitemShut {NoStop}%
\bibitem [{\citenamefont {Fisher}\ \emph {et~al.}(1973)\citenamefont {Fisher},
  \citenamefont {Hornung}, \citenamefont {Brodale},\ and\ \citenamefont
  {Giauque}}]{cmn}%
  \BibitemOpen
  \bibfield  {author} {\bibinfo {author} {\bibfnamefont {R.~A.}\ \bibnamefont
  {Fisher}}, \bibinfo {author} {\bibfnamefont {E.~W.}\ \bibnamefont {Hornung}},
  \bibinfo {author} {\bibfnamefont {G.~E.}\ \bibnamefont {Brodale}}, \ and\
  \bibinfo {author} {\bibfnamefont {W.~F.}\ \bibnamefont {Giauque}},\
  }\bibfield  {title} {\enquote {\bibinfo {title} {Magnetothermodynamics of
  {C}e$_2${M}g$_3$({NO}$_3$)$_{12}\cdot$24{H}$_2${O}. {II}. the evaluation of
  absolute temperature and other thermodynamic properties of cmn to 0.6
  m°{K}},}\ }\href {\doibase 10.1063/1.1679183} {\bibfield  {journal}
  {\bibinfo  {journal} {J. Chem. Phys.}\ }\textbf {\bibinfo {volume} {58}},\
  \bibinfo {pages} {5584--5604} (\bibinfo {year} {1973})}\BibitemShut {NoStop}%
\bibitem [{\citenamefont {Moon}\ \emph {et~al.}(1968)\citenamefont {Moon},
  \citenamefont {Koehler}, \citenamefont {Child},\ and\ \citenamefont
  {Raubenheimer}}]{Yb2O3}%
  \BibitemOpen
  \bibfield  {author} {\bibinfo {author} {\bibfnamefont {R.~M.}\ \bibnamefont
  {Moon}}, \bibinfo {author} {\bibfnamefont {W.~C.}\ \bibnamefont {Koehler}},
  \bibinfo {author} {\bibfnamefont {H.~R.}\ \bibnamefont {Child}}, \ and\
  \bibinfo {author} {\bibfnamefont {L.~J.}\ \bibnamefont {Raubenheimer}},\
  }\bibfield  {title} {\enquote {\bibinfo {title} {Magnetic structures of
  {Er$_2$O$_3$} and {Yb$_2$O$_3$}},}\ }\href {\doibase 10.1103/PhysRev.176.722}
  {\bibfield  {journal} {\bibinfo  {journal} {Phys. Rev.}\ }\textbf {\bibinfo
  {volume} {176}},\ \bibinfo {pages} {722--731} (\bibinfo {year}
  {1968})}\BibitemShut {NoStop}%
\end{thebibliography}
\end{document}